# Simulation of deterministic energy-balance particle agglomeration in turbulent liquid-solid flows

Derrick O. Njobuenwu and Michael Fairweather

School of Chemical and Process Engineering, University of Leeds, Leeds, LS2 9JT, UK

**ABSTRACT**

An efficient technique to simulate turbulent particle-laden flow at high mass loadings within the four-way coupled simulation regime is presented. The technique implements large eddy simulation, discrete particle simulation, a deterministic treatment of inter-particle collisions and an energy-balanced particle agglomeration model. The algorithm to detect inter-particle collisions is such that the computational costs scale linearly with the number of particles present in the computational domain. On detection of a collision, particle agglomeration is tested based on the pre-collision kinetic energy, restitution coefficient and the van der Waals' interactions. The performance of the technique developed is tested by performing parametric studies of the influence the restitution coefficient ($e_n = 0.2$, 0.4, 0.6 and 0.8), particle size ($d_p = 60$, 120, 200 and 316 µm), fluid inertia ($Re_\tau = 150$, 300 and 590) and particle concentration ($\alpha_p = 5.0 \times 10^{-4}$, $1.0 \times 10^{-3}$ and $5.0 \times 10^{-3}$) have on particle-particle interaction events (collision and agglomeration). The results demonstrate that the collision frequency shows a linear dependency on the restitution coefficient, while the agglomeration rate shows an inverse dependence. Collisions among smaller particles are more frequent and efficient in forming agglomerates than those of coarser particles. The particle-particle interaction events show a strong dependency on the shear Reynolds number $Re_\tau$, while increasing the particle concentration effectively enhances particle collision and agglomeration. Overall, the sensitivity of the particle-particle interaction events to the selected simulation parameters is found to influence the population and distribution of the primary particles and agglomerates formed.

**KEYWORDS**: Large eddy simulation, discrete particle simulation, hard-sphere collision, agglomeration, turbulent flow, particle-particle interactions, van der Waals' interaction

I. INTRODUCTION

Of particular interest in this work are the turbulent liquid-solid suspensions encountered in nuclear waste sludge transport and separation processes, where high mass loading is desirable to minimise waste volumes. High particle concentrations with volume fractions over a tenth of a percent result in particle-particle collisions becoming an important physical mechanism [1] in addition to the strong hydrodynamic interactions between the particles and the fluid, although only a fraction of the collisions lead to agglomeration, known as the collision efficiency or agglomeration rate [2, 3]. In the reverse process, agglomerates can also be broken-up into smaller particles due to turbulent shear [4]. In physical space, the agglomeration and break-up processes occur concurrently. Since break-up is a secondary process succeeding the earlier agglomeration process, the rate of break-up will depend on the population and structural dimension of agglomerates in the system. Hence, the balance between agglomeration and break-up events controls the instantaneous particle size distribution. It is therefore important to understand the underlying physics of particle interactions and how they will affect the retrieval and transport of UK legacy nuclear waste sludge required for its efficient and effective clean-up. In addition, particle agglomeration is an ever-present phenomenon in most equipment processing fluids populated with cohesive particle, and in many contexts represents an operational problem.

Undertaking studies on the hydrodynamic transport of suspended particles, taking into account their physicochemical behaviour using physical modelling of particle interactions in turbulence, is a difficult task, even with current technologies. One of the challenges with physical modelling of particle-particle interactions (collision and agglomeration events) is in the difficulties encountered in undertaking well-controlled experiments where individual effects can be isolated from others and investigated in detail. The alternative to physical modelling is the use of computational fluid dynamic (CFD) approaches, which require accurate treatment of the carrier and dispersed phases as well as interphase interactions.

Particle agglomeration is affected by the interplay between two physically independent processes, manifested in three distinct steps [5]: the transport step, governed by the hydrodynamic transport of particles, the collision step, where inter-particle collisions occur, and the adhesion step, governed by the physicochemical interactions between two bodies resulting in the colliding particles either sticking together or bouncing off one another [6, 7]. These steps leading to agglomeration have long been used in the framework of fouling studies. However, the transport, collision and adhesion steps occur at different scales (time and space) and have often been studied separately by researchers with interests in fluid mechanics, hydrodynamics, and the surface and interface sciences.

Two main approaches exist to treating particle-laden turbulent flows, namely the Eulerian-Eulerian approach where both the carrier (liquid) phase and the dispersed (solid) phase are computed in an Eulerian framework, and the Eulerian-Lagrangian approach in which the liquid phase is again calculated on an Eulerian basis and the solid phase is treated as Lagrangian markers [8]. The Eulerian-Eulerian approach is efficient in terms of computational cost but with drawbacks due to the inherent numerical diffusion of the Eulerian treatment of the particles and the lack of physical details of the particle dynamics, inter-particle interactions, and inter-phase interactions in different zones of the flow field. In the Eulerian-Lagrangian approach on the other hand, the particle trajectory model needs fewer assumptions in the formulation of descriptive equations. The trajectories of a large number of particles existing in the flow field are calculated and collision pairs are found from all the trajectories in a deterministic way, which leads to more precise results and more detailed information.

Here, an Eulerian-Lagrangian approach is adopted because it gives detailed information about every particle's position, force and velocity, and since this method allows more flexibility in treating particle agglomeration and agglomerate break-up in a turbulent flow. In this work, therefore, the flow and turbulence of the carrier phase are predicted by the use of numerical resolution of only the largest and most energetic turbulent eddies, with sub-grid scale modelling of the small scales, i.e. large eddy simulation (LES). The motion of the dispersed phase is computed using a discrete particle simulation taking into account all the significant hydrodynamic forces acting on a particle. When particles are in close proximity, they may collide and subsequently agglomerate, and are treated by steps two and three noted above, respectively.

In the second step, there are two common approaches to simulating inter-particle collision depending on whether particle deformation during a collision is explicitly incorporated in the model or not: time-driven simulation (also known as the soft-sphere model or discrete element method) [9-12] and event-driven simulation (also known as the hard-sphere model) [13-18]. Besides the soft- and hard-sphere collision models, a stochastic collision detection model has also been developed and extensively applied by Sommerfeld and co-workers [2, 19]. This paper adopts the hard-sphere collision model which has been applied in the simulation of fluidized beds [14, 20], and in two- and three-phase channel and pipe flows [15-18], to treat the hydrodynamic transport of particles. The third step predicting particle agglomeration, the adhesion step, is achieved by extending the hard-sphere model to handle cohesion or adhesion. Three common extensions and their variants exist for treating the adhesion step [21]: the energy-based agglomeration model [2, 22, 23], the momentum-based model [3, 17, 24-27], and the energy barrier approach based on DLVO interactions [6] (named after the work of Derjaguin and Landau[28], and Verwey and Overbeek[29], which describes the interaction between two particles as an energy barrier determined from the sum of van der Waals' contributions and double-layer electrostatic interactions. This model assumes agglomeration to occur only if the relative kinetic energy of the pair of colliding particles is high enough to overcome the energy barrier. The DLVO interaction model is commonly adopted in aggregation modelling using population balance equations for nano- and colloidal-sized particles in non-flow or in laminar flow regimes, for which a great deal of literature exists (see Henry, et al.[6] and references cited therein).

In the momentum balance model, inter-particle forces are divided into two parts: adhesive forces, including van der Waals' force, the electrostatic force and solid bridge force; and collision forces such as the elastic repulsive force [21]. By comparing the relative value of the adhesive and collision forces, it has been proposed that particles will coalesce if the adhesive forces acting on them are greater than the collision forces, and vice versa [21, 24, 25].

We have adopted the energy-balance agglomeration model, where the particle pre-collision kinetic energy and the dissipated energy due to irreversible deformations and the van der Waals' energy are considered. Particles are considered to bounce off one another after collision if their pre-collision kinetic energy is sufficient to overcome the viscous dissipation arising from the adhesive forces between particles, and coalesce if the opposite holds. The energy-balance agglomeration model has been successfully applied by Sommerfeld and co-workers [2, 19] using flow fields obtained from Reynolds-averaged Navier-Stokes modelling and stochastic-collision models, and applied, for example, to agglomeration in fluidized beds [30]. Jürgens[22] included oblique collisions to the energy based agglomeration model allowing relative tangential velocities at the contact point. Alletto[23] extended the energy based agglomeration model to account for friction at the contact point, allowing the translational and rotational kinetic energy of the particles to be accommodated in the energy balance and the resulting agglomeration hypothesis.

The model adopted and its variants has successfully been applied to study particle agglomeration in laminar and turbulent flows of various complexities [3, 17, 22-25, 31-33]. The process of agglomeration in these studies was considered to be influenced by various parameters: the shear rate, the size of the primary particles, the Hamaker constant, the mechanical properties of the particles and the properties of the carrier phase. The simulations were all based on an extension of the hard sphere particle-particle collision model, originally developed by Matsumoto and Saito[34] and Crowe, et al.[13]. This type of hard-sphere model incorporates a restitution coefficient to account for an inelastic component in the collisional particle deformation, and a Coulombian friction factor to account for sticking and sliding during the collision.

In general the collision and agglomeration statistics of a mono-disperse primary particle depend on the carrier flow characteristics, especially the flow Reynolds number, and the particle size and concentration. Other parameters of importance are the physical and mechanical properties of the particles. From a predictive point of view, the key issues are associated with the accurate prediction of the carrier flow field, the efficient and effective handling of particle-particle collisions, and the physicochemical interactions between colliding particles. Hence, selection of an appropriate CFD technique to ensure the accurate prediction of the flow field characteristics for the length and time scales encountered in practical flows is necessary given the effect its predictions have on particle transport, dispersion and interactions, particularly within complex flow domains.

There is also insufficient literature on the effect of various physical and flow properties of the fluid and particles in simulations of particle transport and interactions in turbulent flow. The present work, therefore, investigates the effect of selected fluid and particle properties that influence inter-particle collision and the probability of a collision leading to agglomeration based on high resolution large eddy simulations and a hard-sphere, energy-based model in a Lagrangian framework using a four-way coupled approach. The simulations are performed in a turbulent channel flow at varying shear Reynolds numbers and at a particle volume fraction that ensures four-way coupling between the fluid and the dispersed phase. In particular, this work examines the sensitivity of particle-particle interactions to selected mechanical and geometric parameters of the particles as well as to the turbulence properties of the flow.

The paper is organised as follows. In Section II, we describe the predictive approaches used in this work covering simulation of the continuous and dispersed phases, together with the methods employed to handle particle-particle interactions and adhesion, and their coupling. In Section III, relevant results obtained in our simulations are given and discussed. Finally, the main findings are summarised and conclusions are drawn in the Section IV.

## II. NUMERICAL METHODOLOGY
### A. Continuous Phase

An LES solver, BOFFIN [35], was used to compute fully-developed turbulent channel flows at shear Reynolds numbers $Re_\tau = u_\tau h/\nu = 150$, 300 and 590, where $u_\tau$ is the shear velocity, $\nu$ is fluid kinematic viscosity and $h$ is the channel half-height. The channel domain was sized as $2\pi \times 2\pi h \times 4\pi h$ which was discretised in physical space with $129 \times 128 \times 128$ grid points, with periodic boundary conditions imposed along the homogenous directions, i.e. the spanwise (y-axis) and streamwise (z-axis), and no-slip conditions at the walls. The LES solver is a finite-volume code, with a co-located grid arrangement of the primary variables, which is based on a fully implicit low-Mach number formulation and is second-order accurate in both space and time. This LES solver been validated thoroughly for many different flows, e.g. [35, 36]

BOFFIN solves the space-filtered mass and momentum conservation equations for an incompressible fluid, with the contributions of the dispersed phase being regarded as point sources of momentum:

$$\frac{\partial \overline{u}_j}{\partial x_j} = 0 \qquad (1)$$

$$\frac{\partial \overline{u}_i}{\partial t} + \overline{u}_j \frac{\partial \overline{u}_i}{\partial x_j} = -\frac{1}{\rho}\frac{\partial \overline{p}}{\partial x_i} + \frac{\partial}{\partial x_j}\overline{\sigma}_{ij} - \frac{\partial}{\partial x_j}\tau_{ij} + \frac{\Pi_i}{\rho} + \frac{\overline{S}_{m,i}}{\rho}, \qquad (2)$$

where $\overline{\sigma}_{ij} = -2\nu \overline{S}_{ij}$ represents the viscous stress, $\overline{S}_{ij} = \frac{1}{2}(\partial \overline{u}_i/\partial x_j + \partial \overline{u}_j/\partial x_i)$ is the filtered strain-rate tensor, $\tau_{ij} = \overline{u_i u_j} - \overline{u}_i \overline{u}_j$ is the sub-grid scale (SGS) tensor which represents the effect of the SGS motions on the resolved motions, $t$ is time, $x_j$ is the spatial co-ordinate directions, $u_j$ is the velocity vector, $p$ is the pressure, and $\rho$ is the density. The SGS tensor is computed using the dynamic version of the Smagorinsky model proposed by Piomelli and Liu[37], with its detailed implementation presented in a recent paper [36]. The filter width Δ is taken as the cube root of the local grid cell volume, $\Delta = (\Delta_x \Delta_y \Delta_z)^{1/3}$. In Eq. (2) $\Pi_3 = -\rho u_\tau^2/h$ is the constant mean pressure imposed along the streamwise direction (z-axis) that drives the flow. $S_{m,i}$ is a source term and accounts for the action on the fluid of the particles, given by the sum of all hydrodynamic forces in the momentum equation of all particles in a fluid computational cell. Further details of the LES approach may be found elsewhere [35, 36]

## B. Dispersed Phase

The motion of a particle in an LES-predicted turbulent flow field can be viewed as a random process, with its position determined by a deterministic part, evaluated in terms of filtered values, and a stochastic component, arising from the SGS turbulent motions of the fluid phase. For a liquid-solid flow, the hydrodynamic forces (drag, shear lift, pressure gradient and added mass) are considered, and a stochastic Markov model [35] is used to represent the influence of the unresolved carrier fluid velocity fluctuations experienced by a stochastic particle over a time interval d$t$ which is added to the deterministic contribution:

$$d\mathbf{v} = \left\{ \frac{(\overline{\mathbf{u}} - \mathbf{v})}{\tau_p} f_D + C_{SL} \frac{3}{4} \frac{\rho}{\rho_p} [(\overline{\mathbf{u}} - \mathbf{v}) \times \overline{\boldsymbol{\omega}}] + \frac{\rho}{\rho_p} \frac{D\overline{\mathbf{u}}}{Dt} + \frac{1}{2} \frac{\rho}{\rho_p} \left( \frac{d\overline{\mathbf{u}}}{dt} - \frac{d\mathbf{v}}{dt} \right) \right\} dt + \left( C_o \frac{k_{sgs}}{\tau_t} \right)^{0.5} d\mathbf{W_t}, \quad (3)$$

$$d\mathbf{x}_p = \mathbf{v}\, dt, \quad (4)$$

where a boldfaced type denotes a matrix-vector and the terms on the right hand side of Eq. (3) are, respectively, contributions from the drag, shear lift, pressure-gradient and added mass forces, and the SGS fluid velocity fluctuations [35]. Gravity and buoyancy forces were not included as the focus was limited to turbulence effects on agglomeration events. The history force was also neglected due to the low Reynolds numbers considered. The particle properties are denoted by the subscript $p$, and fluid properties are either given without subscript (for readability) or by the subscript $f$ (where it enhances clarity). $\mathbf{v}$ and $\mathbf{x_p}$ are the particle instantaneous velocity and position; $\overline{\mathbf{u}}$ and $\overline{\boldsymbol{\omega}} = 0.5(\nabla \times \overline{\mathbf{u}})$ are known resolved fluid velocities and vorticities interpolated at particle position. The term $f_D$ is a non-linear correction due to the particles' finite Reynolds number, $Re_p$, taken from the Schiller and Newman drag correlation, and expressed as $f_D = 1.0 + 0.15 Re_p^{0.687}$. The instantaneous particle Reynolds number is defined as $Re_p = |\bar{\mathbf{u}} - \mathbf{v}| d_p / v$, where $d_p$ is the particle diameter. $\tau_p = \Phi_p d_p^2 / (18v)$ is the particle relaxation time and when normalised by the viscous time-scale $\tau_f = v / u_\tau^2$, gives the particle Stokes number, $\tau_p^+ = \tau_p / \tau_f$, which is then used to characterise the particle response time, with $\Phi_p = \rho_p / \rho$ being the particle to fluid density ratio. Hence, a superscript (+) denotes variables made dimensionless in wall (viscous) units using the fluid kinematic viscosity, $v$, and the fluid shear velocity, $u_\tau$. The shear lift force coefficient $C_{SL}$ accounts for corrections due to small and large particle Reynolds numbers as proposed by Mei[38] and applied by Njobuenwu, et al.[8]. The time derivative $D\mathbf{u}/Dt = \partial \mathbf{u}/\partial t + \mathbf{u} \cdot \nabla \mathbf{u}$ is the fluid acceleration as observed at the instantaneous particle position. The stochastic term (last term in Eq. (3)) consists of a model constant $C_0$ taken as unity and the SGS turbulence kinetic energy, $k_{sgs}$, which accounts for the effects of the SGS stresses on particle dispersion through the use of the Wiener process $d\mathbf{W}_t$. The SGS kinetic energy is obtained using equilibrium arguments from $k_{sgs} = 2\Delta C_s^{2/3} \bar{S}_{ij} \bar{S}_{ij}$, where $\Delta$ is the filter width, $C_S$ the dynamically calibrated Smagorinsky parameter. During a simulation, the increment of the Wiener process, $\Delta \mathbf{W}_t$, is represented by $\Delta W_{t,i} = \xi_i \times \sqrt{\Delta t}$, where $\xi_i$ is a random vector sampled with zero mean and a variance of unity, determined independently for each time step. The interaction between particles and the fluid phase turbulence is taken into account by the time scale $\tau_t = \tau_p$, with other alternative time scales given in Bini and Jones[35].

During the particle motion, particle-turbulence interactions occur in which the particles are dispersed by the turbulence of the continuous phase and the turbulence of that phase is modulated by the presence of the particles. The characteristics of these interactions are accounted for by the momentum exchange between the fluid and the dispersed particles through the momentum term, Eq. (5), which is added as a source term to the fluid Navier-Stokes equation:

$$\bar{S}_{m,i=x,y,z} = -\frac{1}{\Delta^3} \sum_{a=1}^{n_p} \left( m_p \frac{dv_i}{dt} \right). \qquad (5)$$

where $n_p$ is the number of particles present in a particular cell volume and $m_p$ is the mass of each particle in the cell.

## C. Particle-Particle Interactions

The particle-particle interactions involve two stages, namely, the collision stage and the adhesion stage.

### 1. *Deterministic Collision Algorithm*

At the collision stage, the deterministic hard sphere collision model was implemented subject to some assumptions:
- Particles and agglomerates are modelled as spheres, and interaction between the particles is due to binary collisions.
- Collision is frictionless and particle angular momentum is not considered.
- Only small deformations of particles are allowed post-collision.

In modelling the particle binary collisions, the likely collision partners are first identified where, for small time steps, only collisions between neighbouring particles are likely. By using the concept of virtual cells (e.g. [16, 32]) the cost of checking for collisions can be reduced from $O(N_0^2)$, when collisions between all possible particle pairs are considered, to $O(N_0)$ by dynamically adjusting the virtual cell during the simulation to an optimum size such that the number of particles per cell is sufficiently low, with the size user-specified [32]. With high particle volume fractions $\alpha_p > 10^{-3}$ in the system and $N_0 \sim O(10^6 - 10^8)$, the use of virtual cells in reducing the computational cost of checking for collisions from $O(N_0^2)$ to $O(N_0)$ is a significant advantage when adopting this technique for industrial flows. The computational domain is first decomposed into virtual cells. For a near optimal virtual cell size, $d^n$ is dynamically adjusted during the simulation in order to limit the maximum number of particles in the cell, $N_{p,\max}^0$, with this number specified as an input to the CFD code. The optimum value for the factor $d^n$ is given following Alletto[23]:

$$d^n = d^{n-1}\left(N_{p,\max}^0 / N_{p,\max}^{n-1}\right)^{-1/3}, \tag{6}$$

where $d^{n-1}$ is the factor used to decompose the computational domain at the previous time step $(n-1)$, and $N_{p,\max}^0$ and $N_{p,\max}^{n-1}$ are the maximum number of particles allowed to be contained in a virtual cell and the maximum number particles found in one of the virtual cells at the previous time step, respectively. An optimum value for $N_{p,\max}^0$ for particle-particle collisions in a turbulent channel flow has been found to be within the range 10 to 100 [23]. Hence, a value of $N_{p,\max}^0 = 25$ was adopted for the simulations reported in this paper, used to allow a compromise between high accuracy and efficiency in the handling of particle-particle collisions. All particles within the same virtual cell are tagged by the same index. Consequently, collision detection procedures are limited to the particles in each virtual cell.

For a given virtual cell, all collision pairs are first identified using a method similar to that described in [15, 16, 18, 32]. In the interests of completeness, some of the details of the deterministic collision detection, which are available in the references supra, will be repeated. In order for two particles within a virtual cell to collide, two conditions have to be fulfilled. The first condition is that they have to approach each other, i.e. they have to have a negative dot product between their relative velocities, $\mathbf{v}_r = \mathbf{v}_1 - \mathbf{v}_2$, with a relative separation distance, $\mathbf{x}_r = \mathbf{x}_1 - \mathbf{x}_2$. This first condition can be written as:

$$\mathbf{x}_r \cdot \mathbf{v}_{p,r} < 0. \tag{7}$$

The second condition is that the minimum separation distance, $\mathbf{x}_{r,min}$, within a time step has to be less than the sum of the particles' radii. The time $\Delta t_{min}$ at which this minimum separation distance, $\mathbf{x}_{r,min}$, occurs is given by:

$$\Delta t_{min} = -\frac{\mathbf{x}_r \cdot \mathbf{v}_r}{|\mathbf{v}_r|^2}, \tag{8}$$

while the minimum separation is then given as:

$$\mathbf{x}_{r,m} = \mathbf{x}_r + \mathbf{v}_r \Delta t_{min}. \tag{9}$$

Therefore, contact between neighbouring particles within a time step are detected by satisfying the conditions [15]:

$$\left(\Delta t_{min} \leq \Delta t \ \& \ |\mathbf{x}_{r,min}| \leq d_{12}\right) \cup \left(|\mathbf{x}_r| \leq d_{12}\right), \tag{10}$$

where $d_{12} = (d_{p,1} + d_{p,2})/2$. A value of $|\mathbf{x}_{r,min}| = 0$ indicates a head on collision; eccentric collisions correspond to positive values for $|\mathbf{x}_{r,min}|$. When $|\mathbf{x}_{r,min}| = d_{12}$, the particles have just touched one other.

If the conditions in Eq. (10) hold, the time of contact, $\Delta t_c$, needs to be found since $\Delta t_{min}$ can exceed $\Delta t_c$ if the particles are overlapping. The time of contact is obtained from the solution of the relative position vector as a function of time equation, given as:

$$|\mathbf{x}_r + \Delta t_c \mathbf{v}_r|^2 - d_{12} = 0. \tag{11}$$

Equation (11) has two roots, and the best root, which satisfies the condition $\mathbf{x}_r \cdot \mathbf{v}_{p,r} < 0$, is chosen and it is given by:

$$\Delta t_c = \Delta t_{min} + \frac{\mathbf{x_r} \cdot \mathbf{v_r}}{|\mathbf{v_r}|^2}\sqrt{1-K_n}, \qquad (12)$$

where

$$K_n = \frac{|\mathbf{x_r}|^2|\mathbf{v_r}|^2}{(\mathbf{x_r} \cdot \mathbf{v_r})^2}\left(1 - \frac{d_{12}^2}{|\mathbf{x_r}|^2}\right). \qquad (13)$$

The coordinates of the centres of the particles at the instant of collision can be determined from:

$$\mathbf{x_{r,c}} = \mathbf{x_r} + \Delta t_c \mathbf{v_r}. \qquad (14)$$

*2. Energy Balance Agglomeration Model*

The adhesion step succeeds the collision step, subject to the following assumptions:

- In a typical waste sludge [39], the particles are in contact with highly alkaline and high ionic strength salt solutions, where the electrical double layer associated with charged sites on particle surfaces collapses, and electrostatic repulsions that can disperse particles of like charge are inhibited. Hence, only the van der Waals' forces component of the DLVO theory are responsible for post-collision adhesion.
- Agglomeration is based on the pre-collision energy balance and van der Waals' interactions.
- The agglomerate size and structure is based on a volume equivalent sphere.
- Particles are assumed to coalesce if the cohesive force exceeds the segregation force, otherwise they separate.

Agglomeration for the colliding particles is based on an expression which permits agglomeration if the elastic energy (i.e. the relative kinetic energy before the collision minus the dissipated energy) after the compression period of the collision is less than the work required to overcome the van der Waals' forces [23]:

$$\frac{(\mathbf{v_2^-} - \mathbf{v_1^-})^2 - [(\mathbf{v_2^-} - \mathbf{v_1^-})\cdot \mathbf{n_c}]^2(1-e_n^2)}{|(\mathbf{v_2^-} - \mathbf{v_1^-})\cdot \mathbf{n_c}|} \leq \frac{H^*}{6\delta_0^{*2}}\left[(1-e_n^2)\frac{6}{\pi^2 \rho_p^* \overline{\sigma}^*}\frac{d_{p,1}^{*3} + d_{p,2}^{*3}}{d_{p,1}^{*2} d_{p,2}^{*2}(d_{p,1}^* + d_{p,2}^*)}\right]^{1/2}, \qquad (15)$$

where quantities with the superscript * are dimensionless and are defined as: the particle density $\rho_p^* = \rho_p/\rho$, particle diameter $d_p^* = d_p/2h$, Hamaker constant $H^* = H/(\rho u_b^2 2h)$, and yield pressure $\overline{\sigma}^* = \overline{\sigma}/(\rho u_b^2)$; $u_b$ is the bulk velocity and $2h$ is the channel height. Note the superscript (−) denotes quantities before the collision, and the subscripts 1 and 2 denote particles number one and two. The amount of dissipated energy relative to the incident kinetic energy is quantified by $(1-e_n^2)$, in terms of the coefficient of restitution, $e_n$. The restitution coefficient is not constant but depends on factors such as the initial relative approach speed of the particles, their relative velocities of recession after collision and the particle material (see e.g. [40]). The hard-sphere model adopted here, however, treats this coefficient as a constant that can be estimated from empirical investigations. Note that the energy-balance-based agglomeration model, Eq. (15), adopted has been successfully validated against theoretical results for test cases in both laminar and turbulent flow regimes [23-25]. Numerical results obtained using large eddy simulation were found to be in close agreement with theory, and subsequently the energy-based model was applied to investigate the dynamic process of particle agglomeration in vertical fully developed turbulent channel and pipe flows using LES.

When Eq. (15) is not fulfilled (i.e. no agglomeration occurs) and hence the particles bounce apart from each other due to the resulting impulse during impact, the velocities and positions of both particles have to be updated according to Eqs. (16) and (17):

$$\mathbf{v}_1^+ = \mathbf{v}_1^- + \frac{m_{p,2}}{m_{p,1} + m_{p,2}} \left\{ (1+e_n)\left[(\mathbf{v}_2^- - \mathbf{v}_1^-) \cdot \mathbf{n_c}\right]\mathbf{n_c} \right\}$$

$$\mathbf{v}_2^+ = \mathbf{v}_2^- + \frac{m_{p,1}}{m_{p,1} + m_{p,2}} \left\{ (1+e_n)\left[(\mathbf{v}_2^- - \mathbf{v}_1^-) \cdot \mathbf{n_c}\right]\mathbf{n_c} \right\}, \tag{16}$$

$$\mathbf{x}_1^+ = \mathbf{x}_{1,c} + (t_{n+1} - t_c)\mathbf{v}_1^+$$

$$\mathbf{x}_2^+ = \mathbf{x}_{2c} + (t_{n+1} - t_c)\mathbf{v}_2^+, \tag{17}$$

where $\mathbf{x}_{1c}$ and $\mathbf{x}_{2c}$ denote the coordinates of the centre of mass of particles $p_1$ and $p_2$ at the instant of collision, respectively. However, if Eq. (15) holds (i.e. agglomeration occurs) and a volume equivalent agglomerate structure [2, 23, 24] is assumed, as is the absence of interstitial space between agglomerated particles with the same density as the primary particles, then conservation of mass is satisfied. The velocity and position of the centre of mass of the agglomerate, $p_3$, at the end of the current time step is derived based on conservation of the translational momentum of the collision partners [15]:

$$\mathbf{v}_3^+ = \frac{\mathbf{v}_1^- d_{p,1}^3 + \mathbf{v}_2^- d_{p,2}^3}{d_{p,3}^3}, \tag{18}$$

$$\mathbf{x}_3^+ = \frac{\mathbf{x}_{1c} + \mathbf{x}_{2c}}{2} + (t_{n+1} - t_c)\mathbf{v}_3^+. \tag{19}$$

The diameter of the agglomerate, $d_{p,3}$ in Eq. (18), is obtained from conservation of mass as [15]:

$$d_{p,3} = \sqrt[3]{d_{p,1}^3 + d_{p,2}^3}. \tag{20}$$

### D. Coupling Particle Transport, Collision and Agglomeration

Coupling the transport, collision and adhesion steps is not straightforward. The range of the physicochemical interfacial (DLVO) forces is limited to a few tens of nanometres, a distance much smaller than that of typical particle motion and inter-particle distances in fluids. Following an analogy from Henry, et al.[7], a typical energy barrier which a binary collision needs to overcome to effect agglomeration occurs at separation distances as small as $\Delta x = 3$ nm and, in most cases, the eventual energy barrier is found at distances, for collisions with a wall, that are of the order of $\Delta x^+ = 10$ nm. Taking the lowest shear Reynolds number flow considered, $Re_\tau = u_\tau h/\nu = 150$ ($u_\tau = 7.5 \times 10^{-3}$ m s$^{-1}$, $\nu = 10^{-6}$ m$^2$ s), the range of DLVO forces in wall units is then typically of the order of $\Delta x_{DLVO}^+ = \Delta x u_\tau /\nu = 7.5 \times 10^{-3}$. This is much smaller than the typical hydrodynamic distance $\Delta x^+$ which is equal to the distance travelled by particles in one (hydrodynamic) time step: $\Delta t^+ \sim 1$, $\Delta x^+ = u^+ \times \Delta t^+ \sim 1$. In this case, if one were to choose such a time step and introduce DLVO effects as forces in the particle equation of motion, Eq. (3), the particles would leap over the very narrow range where DLVO forces are significant, thus avoiding the particle-surface interaction altogether [7], or causing inter-particle penetration or overlap. Alternatively, the fluid time step could be reduced to the range where DLVO forces are significant. However, doing this will result in very expensive simulations as the time interval for a time step will increase by $O(10^3)$. To circumvent the large-scale differences between distances associated with the actions of the DLVO forces, the size of the particles, the hydrodynamic transport of particles and inter-particle collisions, an efficient modelling approach was adopted such that each step was treated separately during each time step (hydrodynamic transport of particles, inter-particle collision and adhesion of collided particles). Hence, coupling of the transport, collision and adhesion steps was achieved without drastically reducing the overall time step.

## III. RESULTS AND DISCUSSION
### A. Fluid Velocity Fields

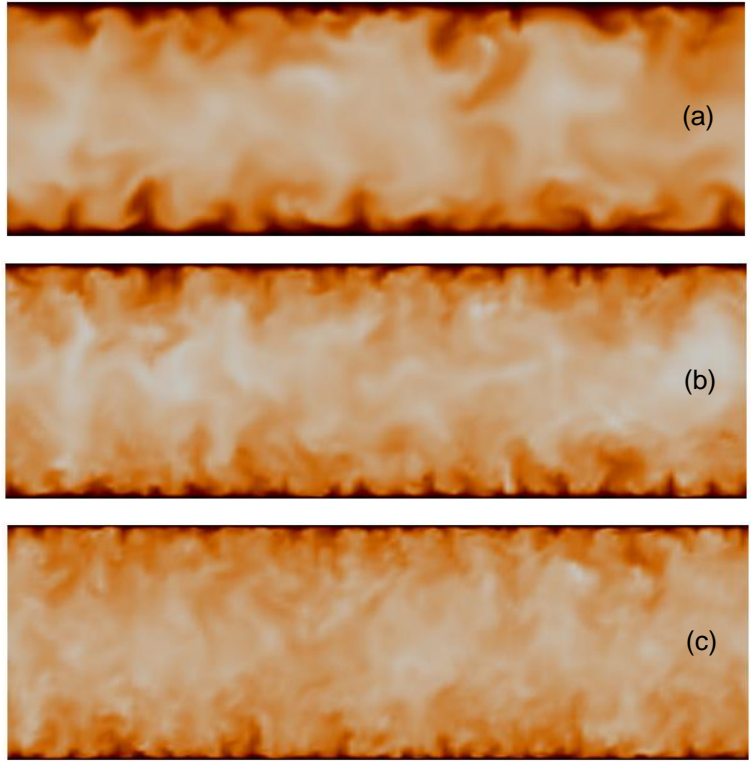

Fig. 1. Instantaneous streamwise velocity field, $w^+$, in $x-y$ planes at $z^+ = 3000$. Contour levels are shown for (a) $0 \leq w^+ < 20$ at $Re_\tau = 150$, (b) $0 \leq w^+ < 21$ at $Re_\tau = 300$, and (c) $0 \leq w^+ < 24$ at $Re_\tau = 590$, from black to white shades.

A visual impression of the change in character of the flow field with increasing shear Reynolds number, $Re_\tau$, is given in Fig. 1, where instantaneous cross-sectional snapshots of the streamwise velocity are shown at one streamwise z-position. Note that the fluid flow field predictions reported in this section were obtained using one-way coupling, i.e. $S_{m,i} = 0$ in Eq. (2), with the particle phase non-existent. The general increase in the range of scales present with increasing Reynolds number is evident across the wall-normal direction, although the large scales dominate for all Reynolds numbers in the central region of the channel. El Khoury, et al.[41] have reported that the average spacing between near-wall low-speed streaks is approximately one half and a tenth of the channel half-height for the lowest and highest $Re_\tau$ flows considered, respectively. This is in conformity with the observed near-wall flow pattern in Fig. 1(a) for the lowest $Re_\tau$ and in Fig. 1(c) for the largest $Re_\tau$.

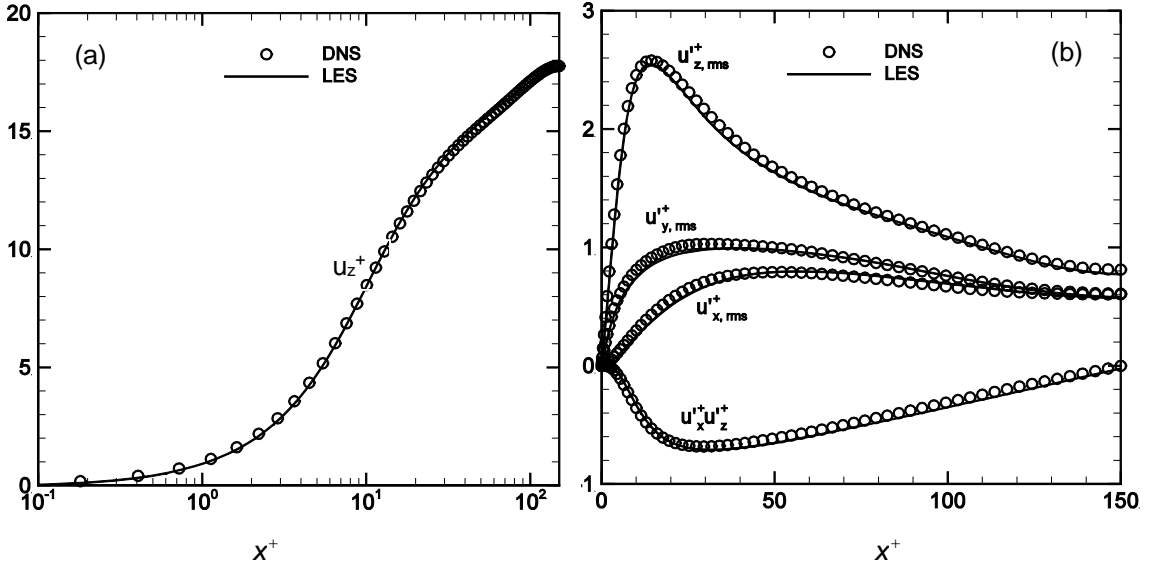

Fig. 2. Statistical moments of the turbulent channel flow: (a) mean streamwise velocity, $(u_z^+)$, and (b) wall-normal $(u'^+_{x,rms})$, spanwise $(u'^+_{y,rms})$ and streamwise $(u'^+_{z,rms})$ root mean square of velocity fluctuations, and Reynolds shear stress $(u'^+_x u'^+_z)$.

Simulation results for the fluid phase were monitored for various averaging start times and averaging periods to evaluate when a statistically stationary state and converged statistics for all shear Reynolds number cases was achieved. Figure 2(a) shows the profile of the mean streamwise velocity, $u_z^+ =<u_z>/u_\tau$, in the wall-normal direction, $x^+ = xu_\tau/\nu$, obtained for one of the three shear Reynolds number flows, $Re_\tau = 150$, with the velocity profiles of the other flows not shown for the sake of brevity. The corresponding profiles of the root mean square of the fluctuating velocity components in the wall-normal, $u'^+_{x,rms} =<u'_{x,rms}>/u_\tau$, the spanwise, $u'^+_{y,rms} =<u'_{y,rms}>/u_\tau$, and the streamwise, $u'^+_{z,rms} =<u'_{z,rms}>/u_\tau$, directions, and of the shear stress $u'^+_x u'^+_z =<u'_x u'_z>/u_\tau^2$, are plotted in Fig. 2(b). The LES results are compared with those obtained from the DNS database of Marchioli, et al.[42] at the same shear Reynolds number. The comparison shows very good agreement, confirming that the use of a highly resolved LES and dynamic modelling of the SGS term gives reliable results. Similar agreement with DNS results was found for the other Reynolds number flows.

## B. Particle-Particle Interactions

The particle equations of motion were integrated using a fourth-order Runge-Kutta scheme, while an interpolation scheme [42] was used to obtain the fluid velocity, $\overline{\mathbf{u}}$, SGS kinetic energy, $k_{sgs}$, and velocity gradients, $\overline{\boldsymbol{\omega}}$, at a particle's position. The particles' initial position, $\mathbf{x}_{p,0}$, at $t^+ = 0$ was random, and the initial velocity was set equal to that of the fluid at the particle's position. For particle-wall interactions, the perfect elastic collision condition was adopted such that all collisions resulted in a rebound back into the computational domain with no loss of kinetic energy. This ideal perfect elastic collision model was adopted to focus attention on the main scope of the work: the deterministic agglomeration model. Periodic boundary conditions were applied along the homogenous directions, making it possible to prolong the duration of the flow by continuous 'recirculation' of the fluid and particles back into the domain.

The computational domain was decomposed into $d^n$ virtual cells in each of the three co-ordinate directions using Eq. (6). For a given virtual cell, all collision pairs were first identified using the method described in §II.C.1 above. Next, these pairs were ordered in ascending order of their time to collision. The first collision in the ordered list (e.g. between the $i^{th}$ and $j^{th}$ particles) for a virtual cell was carried out by advancing those particles to the point of impact by using the velocities at the $n^{th}$ time step. Particle agglomeration was tested based on the agglomeration condition of Eq. (15). The post-collision properties of the colliding particles were changed according to the hard-sphere collision model of Eqs. (16) and (17), or according to the relations shown in Eqs. (18) – (20) if agglomeration was successful. The collision list was then adjusted so that any future collision that contained the $i^{th}$ and $j^{th}$ pair of particles was not allowed to happen in the current time step. By removing the $i^{th}$ and $j^{th}$ particles from future collisions in a given time step, any erroneous multiple collisions of a particle within the same time step were eliminated. The next pair in the list was then allowed to collide, and the process was repeated until there were no more particles left in the collision list. Finally, all particles in the computational domain were advanced for the remainder of the time step using Eqs. (3) and (4).

Using the algorithms developed in this work, we report the performance of the coupled LES and Lagrangian particle tracker with a deterministic particle-particle interaction model. The sensitivity of particle-particle interactions in a turbulent channel flow to four simulation parameters, namely, the shear Reynolds number, particle size, normal restitution coefficient and particle volume fraction, are examined. Three flow shear Reynolds numbers, $Re_\tau = 150$, 300 and 590, four particle sizes, $d_p = 60$, 120, 200 and 316, four normal restitution coefficient values, $e_n = 0.2$, 0.4, 0.6 and 0.8, and three particle volume fractions (concentrations), $\alpha_p = 5.0 \times 10^{-4}$, $1.0 \times 10^{-3}$ and $5.0 \times 10^{-3}$, are examined for a constant density, $\rho_p = 2710$ kg m$^{-3}$. Note that particles were introduced into the computational domain after achieving a fully-developed single-phase flow field, with a new time counter, $t^+ = t u_\tau^2 / \nu = 0$, initiated at this point after which collision and agglomeration counts were taken. It should also be noted that the particle concentration field was still uniformly distributed and had not reach an asymptotic, statistically stationary stage at the start of sampling of the particle-particle interactions. However, in many practical applications, the particle collision statistics shortly after particle release are of more interest than the asymptotic values [43].

### 1. *Dependency of Particle-Particle Interactions on Restitution Coefficient*

First, the sensitivity of particle-particle interactions to the normal restitution coefficient, $e_n$, which denotes the ratio of the relative velocity before and after collision, is examined. The restitution coefficient can be obtained numerically, analytically or measured in laboratory experiments and has been reported [2] as $e_n = 0.4$ for calcite, a nuclear waste sludge simulant. The interest in the sensitivity of the agglomeration process to $e_n$ is necessitated due to the scatter in $e_n$ values reported in the literature and the fact that $e_n$ values for the materials and conditions of interest in this work, required as input to the simulations, are not readily available. The normal coefficient of restitution in the particle-particle interaction model controls how much of the kinetic energy remains for the colliding particles to rebound from one another versus how much is dissipated as heat, or work done in deforming the colliding pair. The amount of dissipated energy relative to the incident kinetic energy is quantified by $(1-e_n^2)$ which appears directly in the agglomeration criterion, Eq. (15), showing that a normal restitution coefficient $e_n = 0$ indicates a complete dissipation of kinetic energy and of relative normal motion, whereas a normal restitution coefficient $e_n > 0$ implies a post-collisional consequence of the particles bouncing off one another or sticking together. For a purely elastic impact with $e_n = 1$, no kinetic energy is dissipated. Hence, an increase in $e_n$ enhances the impulsive force, decreases the amount of energy dissipated during the collision and weakens the cohesive force between the colliding particles, thus reducing the probability of agglomeration conditions being satisfied.

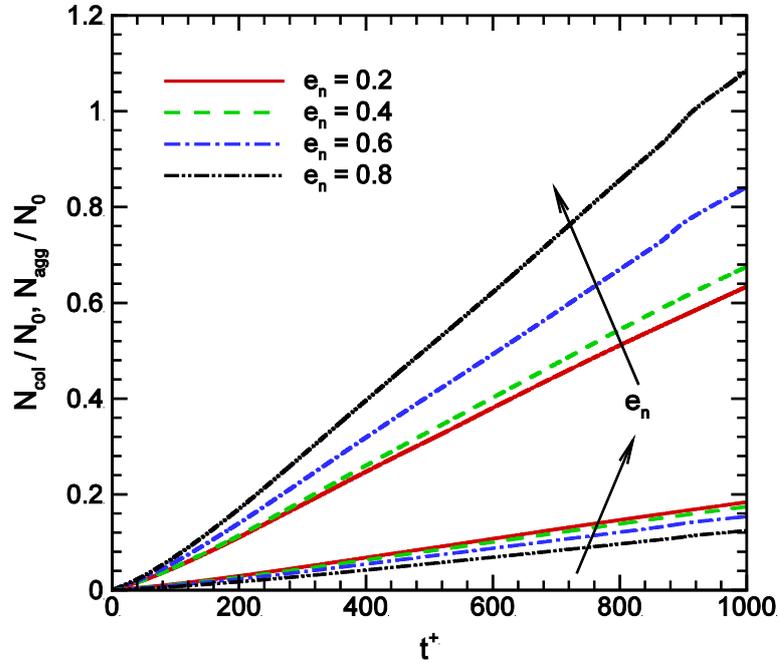

Fig. 3. Distribution of the total number of particle-particle collision events, $N_{col}$, and the total number of particle-particle collisions leading to agglomeration, $N_{agg}$, normalised by the initial total number of primary particles, $N_0$, as a function of the non-dimensional simulation time, $t^+$, for various values of the normal coefficient of restitution, $e_n$ ($Re_\tau = 150$, $d_p = 60$ μm, $\alpha_p = 1 \times 10^{-3}$, upper curves $N_{agg}/N_0$, lower curves $N_{agg}/N_0$).

Figure 3 shows the time evolution of the effect of the normal coefficient of restitution on the number of the accumulated particle-particle collisions, $N_{col}$, and the total number of the accumulated particle-particle collisions resulting in agglomeration, $N_{agg}$, both normalised by the initial total number of primary particles injected, $N_0$, for a shear Reynolds number, $Re_\tau = 150$, primary particle diameter, $d_p = 60$ μm, and particle volume fraction, $\alpha_p = 1 \times 10^{-3}$. Besides the value of $e_n = 0.4$ reported for calcite [2], additional values of $e_n = 0.2$, 0.6 and 0.8 are used to investigate the influence of the coefficient of restitution on particle-particle interactions. Note that unless otherwise stated, sampling of the collision and agglomeration events started at the beginning of the simulation, and the results shown here are for simulations run up to dimensionless time $t^+ = 1000$ in wall units. However, the agglomeration process continued beyond this reported time interval.

From Fig. 3, the number of particle collisions, $N_{col}$, and of agglomerations, $N_{agg}$, varies approximately linearly with time, $t^+$, and shows a strong dependence on the particle normal restitution coefficient. The larger the coefficient of restitution the larger the collision frequency and the smaller the number of such collisions resulting in agglomeration, supporting previous observations [12]. Hence, from Fig. 3 the number of collisions with respect to the normal restitution coefficient at any time has a trend directly opposite to the number of agglomeration processes. This observation slightly differs to that of Breuer and Almohammed[24] who reported that both $N_{col}$ and $N_{agg}$ showed a similar dependency on $e_n$ such that $N_{col}$ and $N_{agg}$ decreased with an increase in $e_n$. However, predictions of collision efficiency from previous works [2, 23, 24] have a similar trend to that of the present predictions, shown in

Fig. 4. The collision efficiency (also known as the agglomeration rate), $N_{agg}/N_{col}$, is defined as the ratio of the total number of accumulated particle-particle collisions leading to agglomeration to the total number of the accumulated particle-particle collisions. Figure 4 shows a sharp change in the profile of $N_{agg}/N_{col}$ from the start of the simulations, during which time the initial conditions are still prevalent. With increases in the simulation time up to $t^+ \sim 100$, more segregations of the particles as well as interactions with turbulent structures occur, and $N_{agg}/N_{col}$ reduces, reaching a constant value as the particle distribution attains a statistically steady state at about $t^+ \sim 200$. Beyond this point, any deviation from steady state is caused by changes in the particle size distribution due to agglomeration, and by further particle segregation under the actions of sweep and eject events in the near-wall region of the channel.

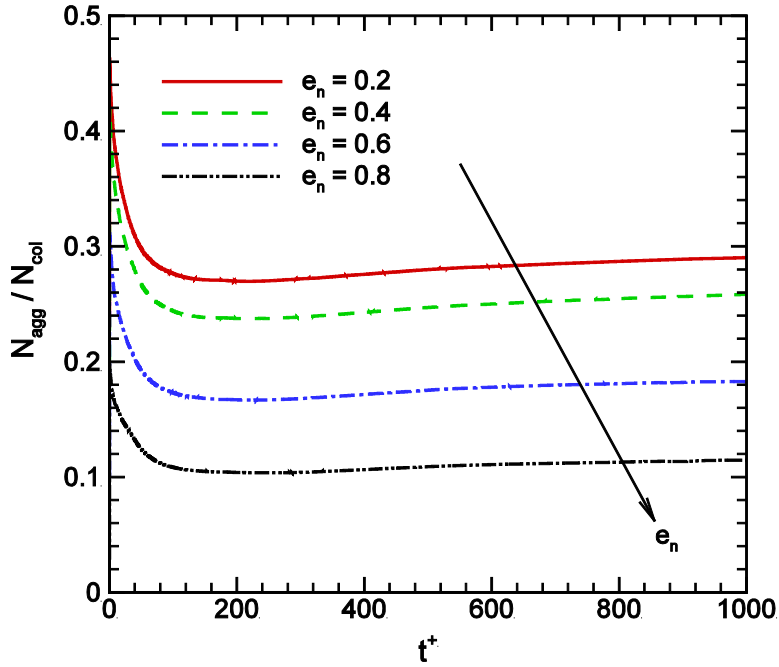

Fig. 4. Distribution of the agglomeration rate, $N_{agg}/N_{col}$, as a function of the non-dimensional simulation time, $t^+$, for various values of the normal coefficient of restitution, $e_n$ ($Re_\tau = 150$, $d_p = 60\,\mu m$, $\alpha_p = 1\times 10^{-3}$).

Additional insights into the influence of the normal restitution coefficient on particle-particle interactions in a turbulent channel flow for the base case ($Re_\tau = 150$, $d_p = 60$ μm, $\alpha_p = 1\times 10^{-3}$) can be obtained from the results of Fig. 5. Figure 5(a) shows the time evolution of the total number of the agglomerated primary particles (i.e. the total number of the primary, or single, particles involved in all agglomerates), $N_{pp}$. As observed by Breuer and Almohammed[24], the results demonstrate that although more agglomerates are formed with time, the number of agglomerated primary particles decreases with an increase in the normal restitution coefficient. This observation of $N_{pp}$ decreasing with an increase in $e_n$ is expected as fewer agglomerates are formed at higher $e_n$, as discussed in relation to Figs. 3 and 4. The total number of agglomerates (multiple particles, excluding single particles), $N_a$, irrespective of the agglomerate type (or size), over time is shown in Fig. 5(b). Similar to the observations from Fig. 5(a), the total number of agglomerates formed increases with time and is inversely proportional to the value of the normal restitution coefficient.

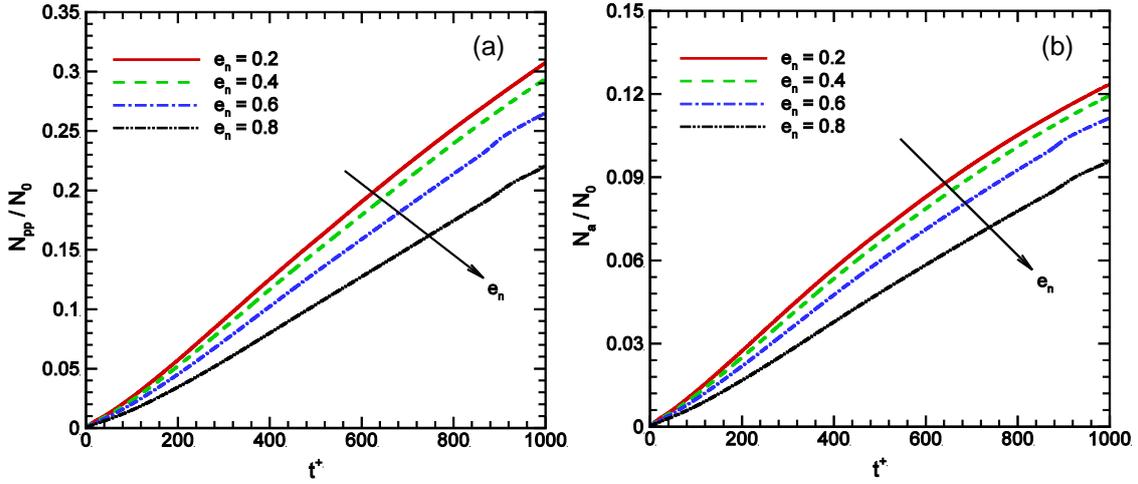

Fig. 5. Influence of the normal restitution coefficient, $e_n$, for particle-particle interaction on the time evolution of: (a) the total number of the agglomerated primary particles, $N_{pp}$, and (b) the total number of agglomerates, $N_a$, both normalised by the initial total number of primary particles, $N_0$ ($Re_\tau = 150$, $d_p = 60\,\mu m$, $\alpha_p = 1\times 10^{-3}$).

Figure 6(a) shows the average number of primary particles included in an agglomerate, defined as the ratio of the total number of agglomerated primary particles, $N_{pp}$, to the total number of agglomerates, $N_a$, and its variation with dimensionless time, $t^+$. These results again illustrate that the size of the agglomerates with respect to the number of primary particles included in the agglomerate increases with a decrease in the normal restitution coefficient. This is again expected and, with respect to Figs. 3 to 5, a large $e_n$ favours a post-collisional state of particles bouncing off one another whilst a small $e_n$ encourages particles involved in binary collisions to stick together. Within the time interval reported, up to $t^+ \sim 1000$, the average number of primary particles included in an agglomerate, $N_{pp}/N_a$, falls within the double and triple particle agglomerate range as $N_{pp}/N_a \sim 2-3$. From the onset of agglomerates being formed from a population of single particles, the value of $N_{pp}/N_a$ is about 2 as the double particles are the first agglomerate size to be formed. As the simulation time progresses, most of the double particles already formed combine with other agglomerates or singles to form agglomerates of higher size, hence, the average value $N_{pp}/N_a$ shifts away from 2 towards 3, as is evident in Fig. 6(a). This is because the lesser multiple particle agglomerates are more readily formed than larger agglomerates. The precursors for the formation of two and three particle agglomerates are the single particles, and they have the largest number density within the time interval reported, as will be shown later in Fig. 7. Furthermore, and again as will be shown later, collision and agglomeration processes are strongly dependent on the size of the two particles involved in any particle-particle interaction. More collisions and subsequent agglomerations therefore take place the smaller the particle size. Figure 6(b) compares the number of primary particles (i.e. single (1)) and of agglomerates of the same type (double (2), triple (3), quadruple (4), quintuple (5), sextuple (6), etc. particles) formed after a simulation time $t^+ = 1000$ for the four normal restitution coefficients ($e_n = 0.2$, 0.4, 0.6 and 0.8) investigated. For the agglomerate sizes reported, an increase in $e_n$ produces a larger population of single particles, but with a reduction, in general, in the number of agglomerates of the same size. The size of the agglomerates in terms of the number of primary particles included in the agglomerate, $N_{pp}$, generally increases with a decrease in the normal restitution coefficient, particularly for agglomerates with up to ten primary particles. Further simulation time would allow agglomerates made up of more than 10 primary particles to be considered, although clear trends can be established over the total simulation time considered, making longer run times unnecessary.

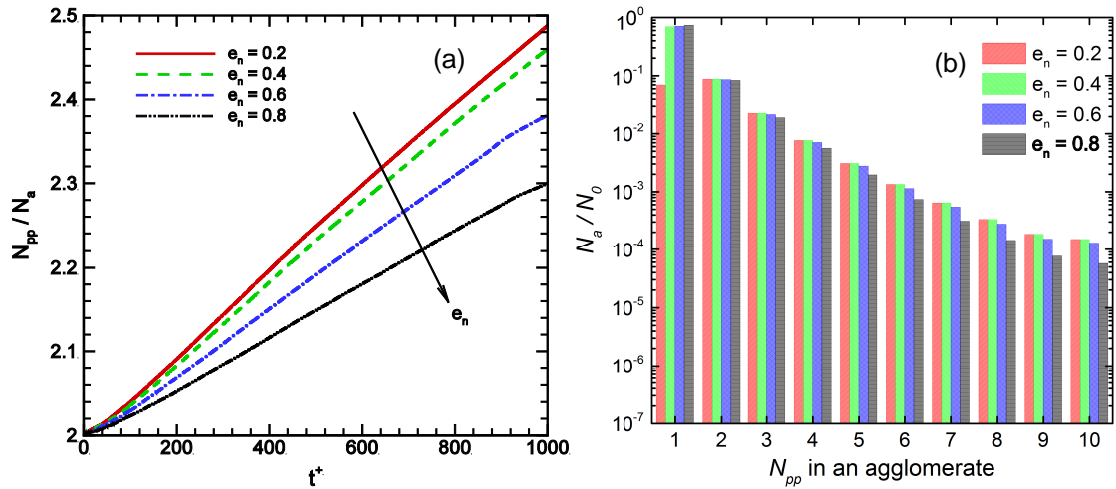

Fig. 6. Influence of the normal restitution coefficient, $e_n$, for particle-particle interaction on: (a) the time evolution of the average number of primary particles included in an agglomerate, $N_{pp}/N_a$, and (b) the number of particles or agglomerates of the same type (single (1), double (2), triple (3), quadruple (4), etc. particles) after a simulation time, $t^+ = 1000$ ($Re_\tau = 150$, $d_p = 60\,\mu$m, $\alpha_p = 1\times 10^{-3}$).

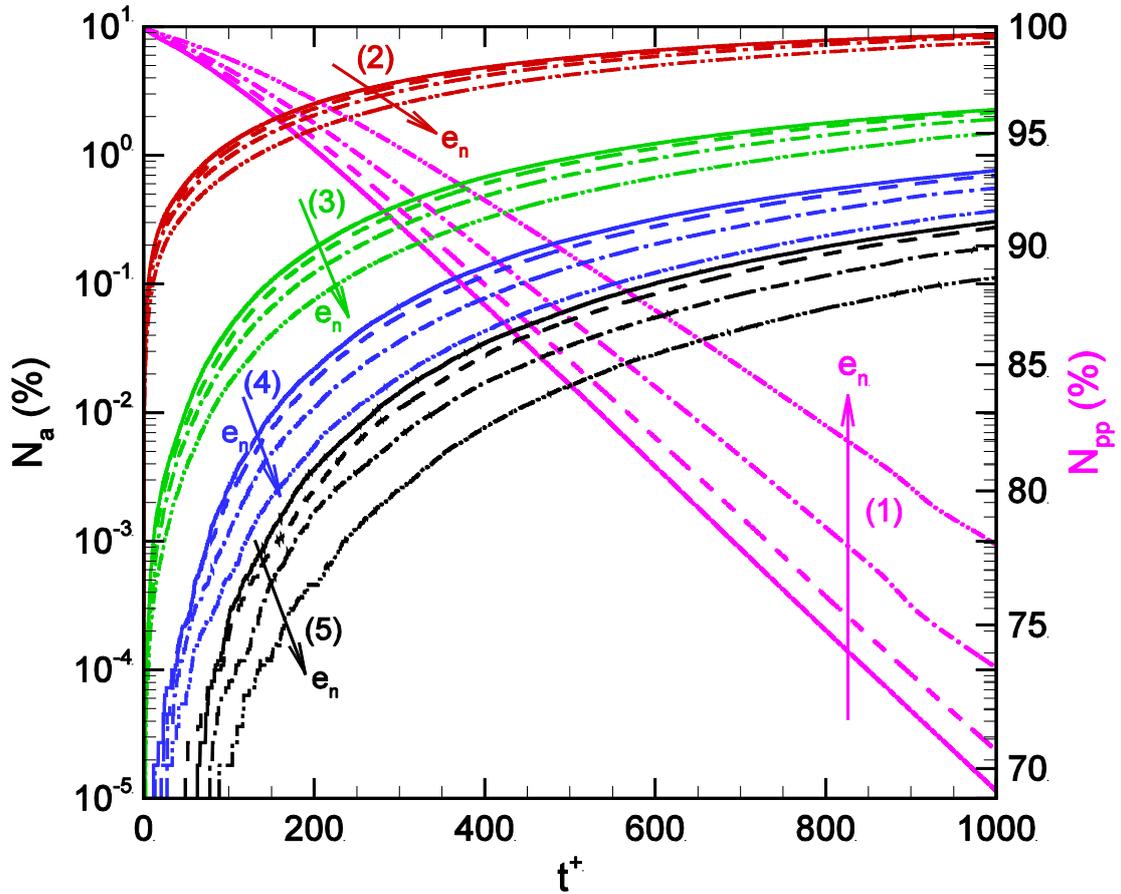

Fig. 7. Influence of the normal restitution coefficient, $e_n$, for particle-particle interaction on time evolution of the population of single and multiple particles. Line numbers: single (1), double (2), triple (3), quadruple (4), quintuple (5) ($Re_\tau = 150$, $d_p = 60\,\mu m$, $\alpha_p = 1\times 10^{-3}$).

Fig. 7 shows the population balance (growth and/or death) of the single and multi-sized particle agglomerates, defined as the percentage of the ratio of the number of particle sizes to the initial number of total single particles, after a simulation time $t^+ = 1000$. The results of Fig. 7 complement the earlier observations noted in regards to Fig. 6. Figure 7 therefore demonstrates that an increase in $e_n$ decreases the rate of depletion of the primary particles, such that at any given time the number of single particles in the system increases and the size of the various agglomerates, in terms of the number of primary particles included in the agglomerate, decreases with increasing $e_n$, as was observed by Breuer and Almohammed[24].

Table 1. Values of particle-particle collisions ($N_{col}/N_0$), agglomeration events ($N_{agg}/N_0$), agglomeration rate ($N_{agg}/N_{col}$), agglomerated primary particles ($N_{app}/N_0$), agglomerate number ($N_a/N_0$) and average number of primary particles included in an agglomerate ($N_{pp}/N_a$) for different normal restitution coefficients, $e_n$, after a dimensionless time $t^+ = 1000$ ($Re_\tau = 150$, $d_p = 60\,\mu m$, $\alpha_p = 1\times 10^{-3}$).

| $e_n$ | $N_{col}/N_0$ | $N_{agg}/N_0$ | $N_{agg}/N_{col}$ | $N_{pp}/N_0$ | $N_a/N_0$ | $N_{pp}/N_a$ |
|---|---|---|---|---|---|---|
| 0.2 | 0.634 | 0.184 | 0.290 | 0.307 | 0.124 | 2.488 |
| 0.4 | 0.676 | 0.175 | 0.258 | 0.294 | 0.119 | 2.460 |
| 0.6 | 0.842 | 0.154 | 0.183 | 0.265 | 0.111 | 2.382 |
| 0.8 | 1.086 | 0.125 | 0.115 | 0.220 | 0.096 | 2.300 |

Overall, the notable feature in Figs. 3 to 7 is the strong dependence of particle-particle interactions on the coefficient of restitution, $e_n$, with a summary of key simulation parameters at dimensionless time $t^+ = 1000$ given in Table 1. Increasing $e_n$ from 0.2 to 0.8 has a dramatic effect on the parameters reported, such that $N_{col}/N_0$ increases from 0.634 to 1.086 showing a linear proportionality, $N_{agg}/N_0$ diminishes from 0.184 to 0.125 implying an inverse relationship, and $N_{agg}/N_{col}$ decreases from 0.290 to 0.115, again suggesting an inverse dependence. Others parameters such as $N_{pp}/N_0$ reduce from 0.307 to 0.220, $N_a/N_0$ decreases in value from 0.124 to 0.096, while $N_{pp}/N_a$ reduces from 2.488 to 2.300, demonstrating that agglomeration-related parameters generally depend strongly on $e_n$ in an inverse fashion. One obvious feature of the results is therefore that the larger the coefficient of restitution, $e_n$, the larger the number of collision events, $N_{col}$, and the smaller the value of the agglomeration events, $N_{agg}$, $N_{agg}/N_{col}$, $N_a$ and $N_{pp}$.

## 2. *Effect of Particle Size on Particle-Particle Interactions*

Particle size has an influence on the value of the two forces on both sides of the agglomeration equation, Eq. (15). Particle size (or inertia) has therefore been shown to have a significant influence on the relative velocity between the two particles. Agglomeration occurs when the attractive forces are predominant; and the smaller the particle size is, the higher these forces are, as is evident in Eq. (15). Hence, agglomeration processes are sensitive to the size of the primary particles. Lin and Wey[44] reported that the tendency for particles to agglomerate is proportional to the surface area of the particles. Small particles have a larger surface area per unit volume, thereby favouring agglomeration.

In order to investigate the effect of particle size on particle-particle interaction processes, it is necessary that the global particle volume fraction, $\alpha_p$, and not the number of primary particles, $N_0$, be comparable. $N_0$ values used in the simulations were 75,250 for $d_p = 316$ µm particles, increasing to 10,992,290 for $d_p = 60$ µm particles. Since the $N_0$ values used for the four particle sizes ($d_p = 60$, 120, 200 and 316 µm) simulated are not comparable, the particle-particle interaction parameters have to be normalised by $N_0$ for comparisons to be based on a unit particle. The four particle sizes examined can alternatively be organised according to the values of their Stokes number, $\tau_p^+$. The equivalent Stokes numbers of the four particle sizes are $\tau_p^+ = 3.62 \times 10^{-2}$, $1.45 \times 10^{-1}$, $4.02 \times 10^{-1}$ and 1.0, respectively. Figure 8 addresses the effects of particle size $d_p$ (or inertia $\tau_p^+$) on collision and agglomeration behaviour in a turbulent channel flow. Figures 8(a), (b) and (c) show semi-logarithmic plots of the evolution of the normalised total number of particle-particle collision events, $N_{col}/N_0$, the normalised total number of particle-particle collisions leading to agglomeration, $N_{agg}/N_0$, and the agglomeration rate ($N_{agg}/N_{col}$), respectively, as a function of the size of the primary particles, $d_p$. For these simulations a normal restitution coefficient $e_n = 0.4$ for calcite, a shear Reynolds number $Re_\tau = 150$ and an initial global volume fraction $\alpha_p \sim 1 \times 10^{-3}$ were used.

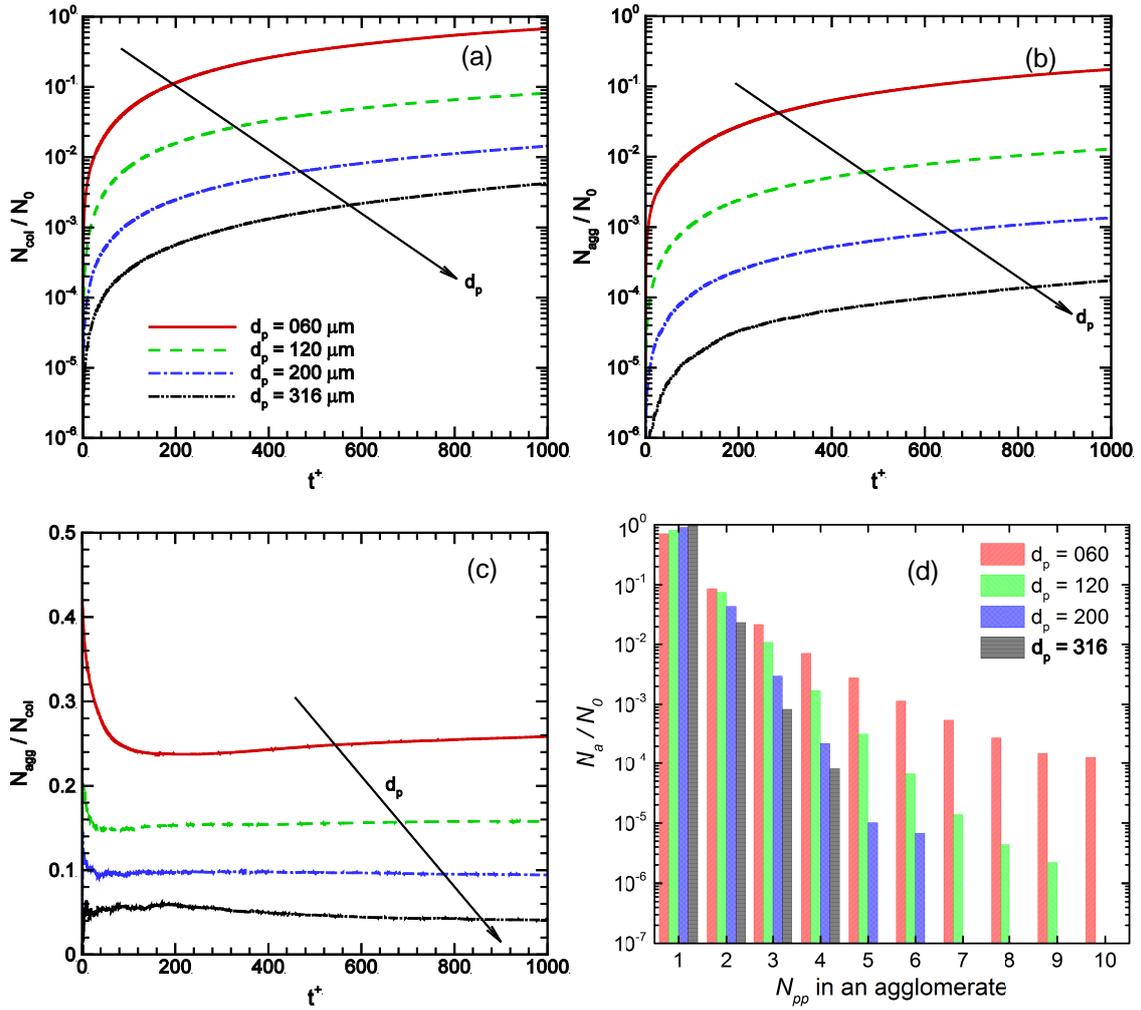

Fig. 8. Distribution of: (a) total number of particle-particle collision events, $N_{col}$, (b) total number of particle-particle collisions leading to agglomeration, $N_{agg}$, both normalised by initial total number of primary particles, $N_0$, (c) agglomeration rate, $N_{agg}/N_{col}$, all as a function of non-dimensional simulation time, $t^+$, and (d) number of particles or agglomerates, $N_a/N_0$, of the same type (single (1), double (2), triple (3), quadruple (4), etc. particles) after a simulation time, $t^+ = 1000$ ($d_p = 60$, 120, 200 and 316 µm, $Re_\tau = 150$ $e_n = 0.4$ ;, $\alpha_p = 1\times 10^{-3}$).

It is clearly demonstrated, from the results presented in Figs. 8(a-c), that the collision frequency, $N_{col}/N_0$, the collision efficiency, $N_{agg}/N_0$, and the agglomeration rate, $N_{agg}/N_{col}$, all decrease with an increase in particle size, $d_p$. These results are in qualitative agreement with theory and with observations reported elsewhere [2, 21, 24, 45, 46]. According to theory [21, 46], the kinetic energy between colliding coarse particles is high, whilst that between fine and coarse particles is slightly less, and that between fine particles is the lowest. In contrast to the kinetic energy between particles, the adhesion energy between small particles is larger than that between coarse particles. However, the sticking behaviour of two individual particles depends on the relative value of the adhesion energy and the kinetic energy between them, as represented in the agglomeration model, Eq. (15). Particles rebound if the kinetic energy is higher than the adhesion energy, while they coalesce if the inverse is true. Reddy and Mahapatra[45] observed that agglomeration in fluidized bed combustion power plant takes place when the coal material contains either too many fine particles or coarse particles, or both in very large proportions. The results of Wang, et al.[21] complemented the latter authors' observations, finding that compared with coarse particles, small particles lead to agglomeration much more readily, and the coalescence of small particles, or small and coarse particles, is the main mechanism by which agglomeration occurs. Numerical experiments based on full DLVO theory agree with the present results on the sensitivity of particle-particle interactions to particle size. DLVO theory therefore predicts a marked increase of the total interaction energy with an increase in particle size, and therefore a dramatic decrease in the rate of aggregation of colloidal particles [6]. Hence, the high number of inter-particle collisions, $N_{col}/N_0$, with reducing particle size, $d_p$, in Fig. 8(a) is a prerequisite for the large number of agglomeration processes, $N_{agg}/N_0$, in Fig. 8(b), assuming that the cohesive force is large enough with reducing particle size. The influence of particle size on the number of agglomerates of the same type (double, triple, quadruple, etc.) and the number of non-agglomerated primary particles (single) is shown in Fig. 8(d). After the referenced simulation time, $t^+ \sim 1000$, the smallest particles with $d_p = 60\,\mu m$ have undergone more agglomeration processes than the other sizes of particle, hence, there is a smaller number of single particles present in the computational domain, followed by the $d_p = 120$, 200 and 316 μm particle sizes. Within this time frame, the number agglomerates, $N_a/N_0$, of the same type decreases with an increase in the particle size. This observation complements the earlier finding from the results given in Fig. 8(a-c) that the rate of inter-particle collision and agglomeration decreases as the particle size increases. Hence, only primary particle sizes $d_p = 60$ and 120 μm formed agglomerates with a primary particle number $N_{pp}$ beyond six, whilst only the $d_p = 60$ μm primary particles generated agglomerates with up to 10 constituents.

This relationship between particle-particle interaction and particle size, summarised in Table 2, is as earlier stated and is generally ascribed to the fact that smaller particles have an increased total surface area which leads to more contact, and a higher probability of collision, between particles as compared to their coarser counterparts. As the primary particle size, $d_p$, increases from 60 to 316 µm, the simulated number of inter-particle collisions, $N_{col}/N_0$, decreases from 0.675 to 0.532, the number of agglomeration processes, $N_{agg}/N_0$, reduces from 0.175 to 0.025, and the agglomeration rate, $N_{agg}/N_{col}$, diminishes from 0.258 to 0.041 by the end of the simulations at time $t^+ = 1000$, all showing an inverse proportionality dependence. It is therefore reasonable to infer that smaller particles promote inter-particle collisions and the formation of agglomerates.

Table 2. Values of particle-particle collisions ($N_{col}/N_0$), agglomeration events ($N_{agg}/N_0$), agglomeration rate ($N_{agg}/N_{col}$), agglomerated primary particles ($N_{app}/N_0$), agglomerate number ($N_a/N_0$) and average number of primary particles included in an agglomerate ($N_{pp}/N_a$) for different particle sizes after a dimensionless time $t^+ = 1000$ ($Re_\tau = 150$, $e_n = 0.4$, $\alpha_p = 1\times10^{-3}$).

| $d_p$/(µm) | $\tau_p^+$ | $N_{col}/N_0$ | $N_{agg}/N_0$ | $N_{agg}/N_{col}$ | $N_{pp}/N_0$ | $N_a/N_0$ | $N_{pp}/N_a$ |
|---|---|---|---|---|---|---|---|
| 60 | 3.62×10⁻² | 0.676 | 0.175 | 0.258 | 0.294 | 0.119 | 2.460 |
| 120 | 1.45×10⁻¹ | 0.654 | 0.103 | 0.158 | 0.191 | 0.088 | 2.177 |
| 200 | 4.02×10⁻¹ | 0.532 | 0.050 | 0.094 | 0.097 | 0.047 | 2.074 |
| 316 | 1.0 | 0.620 | 0.025 | 0.041 | 0.050 | 0.024 | 2.040 |

### *3. Effect of Shear Reynolds Number on Particle-Particle Interactions*

Another important parameter that influences particle-particle interactions is the fluid inertia, measured by the flow shear Reynolds number, $Re_\tau$. Note that the effect of different flow Reynolds numbers on particle-particle interactions is equivalent to investigating the impact of turbulence, or fluid velocity, on the same particle-particle interactions. Hence, with other parameters being invariant, the fluid, and by extension the particle, velocity both have a linear relationship with the flow Reynolds number. It is also important to further note that turbulence-induced agglomeration is commonly categorised into two mechanisms, namely, turbulent fluctuations that cause relative motion between particles (the turbulent transport effect) and the preferential concentration that leads to a highly intermittent local pair density distribution and thus an additional enhancement of the average collision rate (the accumulation effect) [43]. The effects of both mechanisms on particle-particle interactions were considered as a whole and not individually.

The sensitivity of particle-particle interactions to the flow shear Reynolds number for a given fixed particle Stokes number, $\tau_p^+ = 1.0$, is shown in Fig. 9. A change in the flow Reynolds number alters the fluid velocities seen by the particles as well as the particles' velocities. With an increase in flow velocity, the ratio of the particle relative kinetic energy to that of the cohesive force is also affected. For a fixed particle inertia, measured by its Stokes number, the number of inter-particle collisions in Fig. 9(a), and the number of collisions leading to agglomeration in Fig. 9(b), both show a strong sensitivity to the flow shear Reynolds number. On the other hand, the agglomeration rate in Fig. 9(c) shows a weak dependence on the flow inertia. Increasing the flow shear Reynolds number shows an increase in the normalised number of particle-particle collision, $N_{col}/N_0$, in Fig. 9(a) and an increase in the normalised number of collisions leading to agglomeration, $N_{agg}/N_0$, in Fig. 9(b). However, the agglomeration rate, $N_{agg}/N_{col}$, in Fig. 9(c) shows an inverse proportionality with a stronger dependence on the shear Reynolds number at early simulation times. The relationship between $N_{agg}/N_{col}$ and $Re_\tau$ becomes weaker as time progresses due to the particle concentration becoming more developed. The relative number of single and multiple particle agglomerates present in the computational domain at the end of the simulation at $t^+ = 1000$ is shown in Fig. 9(d) when the primary particle Stokes number is fixed at $\tau_p^+ = 1.0$. This demonstrates (although difficult to see from the plot) that the relative number of single particles remaining at the end of the simulation is reduced as the shear Reynolds number is increased from $Re_\tau = 150$ to $Re_\tau = 590$. This means that the ratio of the total number of single particles involved in forming agglomerates to the initial number of single particles input to the computational domain, $N_a/N_0$, is smallest for the lowest shear Reynolds number case. This behaviour corroborates that found in the results of Fig. 9(b) and (c) where fewer collisions result in agglomeration the lower the shear Reynolds number. In terms of the agglomerates, Fig. 9(d) shows that an increasing number of double and triple particle agglomerates occur as the shear Reynolds number increases from $Re_\tau = 150$ to 590. Beyond the triple particle agglomerates, no clear relationship between the agglomerate particle number and $Re_\tau$ is evident, with the maximum agglomerate size formed at the end of the simulation consisting of nine primary particles for the $Re_\tau = 590$ case. Agglomerates of five particles and beyond were not found for the lowest shear Reynolds number case, whilst only agglomerates consisting of up to five particles were observed for the moderate shear Reynolds number, $Re_\tau = 300$.

Table 3 shows that at the end of the investigated simulation time $t^+ = 1000$, as the flow shear Reynolds number $Re_\tau$ increases from 150 to 590, for a fixed primary particle Stokes number $\tau_p^+ = 1$, the normalised number of inter-particle collisions, $N_{col}/N_0$, increases from 0.620 to 0.958, and the normalised number of agglomeration processes, $N_{agg}/N_0$, increases from 0.025 to 0.038, again showing an inverse proportionality dependence. In contrast, the agglomeration rate, $N_{agg}/N_{col}$, remains roughly constant over all $Re_\tau$. It can therefore be inferred that the Reynolds number has a minimal effect on the formation of agglomerates.

Table 3. Values of particle-particle collisions ($N_{col}/N_0$), agglomeration events ($N_{agg}/N_0$), agglomeration rate ($N_{agg}/N_{col}$), agglomerated primary particles ($N_{app}/N_0$), agglomerate number ($N_a/N_0$) and average number of primary particles included in an agglomerate ($N_{pp}/N_a$) for different shear Reynolds number $Re_\tau$ after a dimensionless time $t^+ = 1000$ ($\tau_p^+ = 1$, $e_n = 0.4$, $\alpha_p = 1\times10^{-3}$).

| $Re_\tau$ | $d_p$/(μm) | $N_{col}/N_0$ | $N_{agg}/N_0$ | $N_{agg}/N_{col}$ | $N_{pp}/N_0$ | $N_a/N_0$ | $N_{pp}/N_a$ |
|---|---|---|---|---|---|---|---|
| 150 | 316 | 0.620 | 0.025 | 0.041 | 0.050 | 0.024 | 2.040 |
| 300 | 158 | 0.730 | 0.032 | 0.044 | 0.063 | 0.031 | 2.044 |
| 590 | 80.25 | 0.958 | 0.038 | 0.040 | 0.074 | 0.036 | 2.074 |

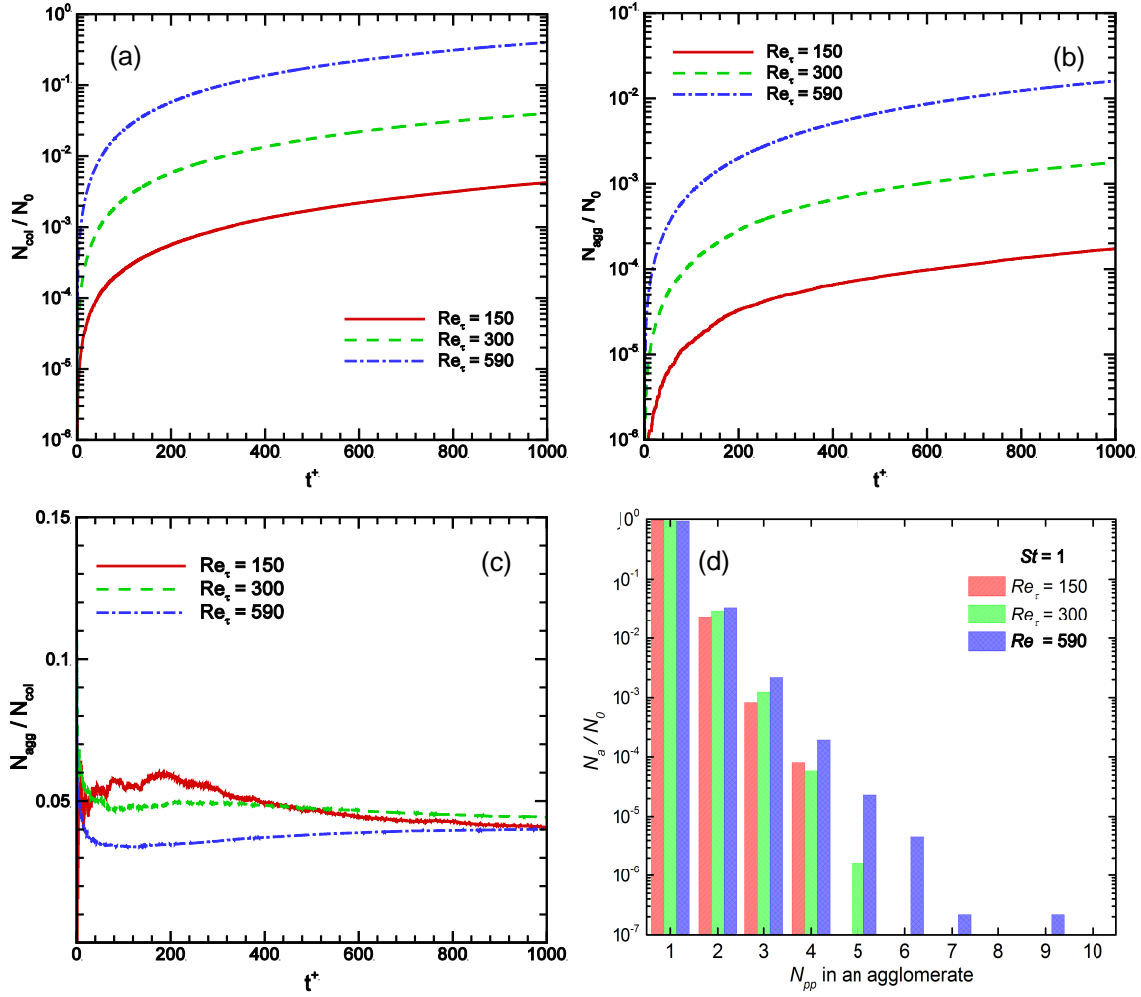

Fig. 9. Distribution of: (a) total number of particle-particle collision events, $N_{col}$, (b) total number of particle-particle collisions leading to agglomeration, $N_{agg}$, both normalised by initial total number of primary particles, $N_0$, (c) agglomeration rate, $N_{agg}/N_{col}$ all as a function of non-dimensional simulation time, $t^+$, and (d) number of particles or agglomerates, $N_a/N_0$, of the same type (single (1), double (2), triple (3), quadruple (4), etc. particles) after a simulation time $t^+ = 1000$ ($Re_\tau = 150$, 300 and 590, particle Stokes number $\tau_p^+ = 1$, $e_n = 0.4$, $\alpha_p = 1 \times 10^{-3}$).

## 4. Correlation Between Degree of Particle-Particle Interactions and Particle Concentration

The results of Fig. 10(a and b) suggest that there exists a strong correlation between the numbers of particle-particle collisions, $N_{col}/N_0$, and agglomerations, $N_{agg}/N_0$, and the particle volume fraction, $\alpha_p$, while in Fig. 10(c) a weak relationship between the agglomeration rate, $N_{agg}/N_{col}$, and the particle volume fraction is exhibited. As shown in Fig. 10(a), the normalised accumulated number of particle collisions continuously increases with increasing simulation time, $t^+$, while the number of collisions increases with the particle volume fraction. This trend is consistent with theory as well as with the observations of Ernst and Sommerfeld[47] where the average distance between particles and computed collision times decreased with increasing volume fraction, consequently increasing the probability of the particles being close enough to cause collisions. As for the agglomeration rate, $N_{agg}/N_{col}$, it can be seen from Fig. 10(c) that its particle volume fraction dependence is weak as the difference in the agglomerate rate with time is similar both in trend and in magnitude, irrespective of the volume fraction. After an initial instability in the value of $N_{agg}/N_{col}$ due to the evolving particle concentration distribution, the agglomeration rate shows a roughly constant profile with time, with the case with the largest volume fraction, $\alpha_p = 5 \times 10^{-3}$, having the highest agglomerate rate at all times reported. The two cases with lower volume fractions, $\alpha_p = 5 \times 10^{-4}$ and $\alpha_p = 1 \times 10^{-3}$, show no significant differences in the magnitude of the agglomeration rate after the initial settling period. Overall, the profiles shown in Fig. 10(a-c) demonstrate that the rate of normalised particle collisions, in Fig. 10(a), and the normalised agglomeration rate, in Fig. 10(b), are similar in terms of their ratio, as given in Fig. 10(c). As expected, the results of Fig. 10(d) indicate that the depletion of the primary single particles at the end of the simulation, $t^+ = 1000$, is highest for the very concentrated case, $\alpha_p = 5 \times 10^{-3}$, with this depletion decreasing with particle volume fraction. The same reasoning given in relation to the results of Fig. 10(a-b) applies here. In terms of the number of multiple particle agglomerates present at the end of the simulation, a linear dependency is evident between the number of agglomerates of a particular type and the volume fraction. The higher the concentration the more agglomeration occurs, and the larger the number of agglomerates of a particular type are present in the computational domain. Hence, the case with the highest particle volume fraction, $\alpha_p = 5 \times 10^{-3}$, shows the largest number of agglomerates of all types, followed by the moderate and then lowest concentration cases. Only agglomerates of up to six and four particles are formed at the end of the simulation for cases with particle volume fractions of $\alpha_p = 1 \times 10^{-3}$ and $\alpha_p = 5 \times 10^{-4}$, respectively, due to the slower collision and agglomeration rates in these cases.

Lastly, Table 4 gives numerical values of the various collision and agglomeration parameters considered in Fig. 10 at the end of the simulation for the case involving a shear Reynolds number, $Re_\tau = 150$, particle normal restitution coefficient, $e_n = 0.4$, and primary particle size, $d_p = 60\,\mu m$. Here, the correlation between the degree of particle-particle interactions and agglomerations, and the particle concentration, as noted above, is again clear.

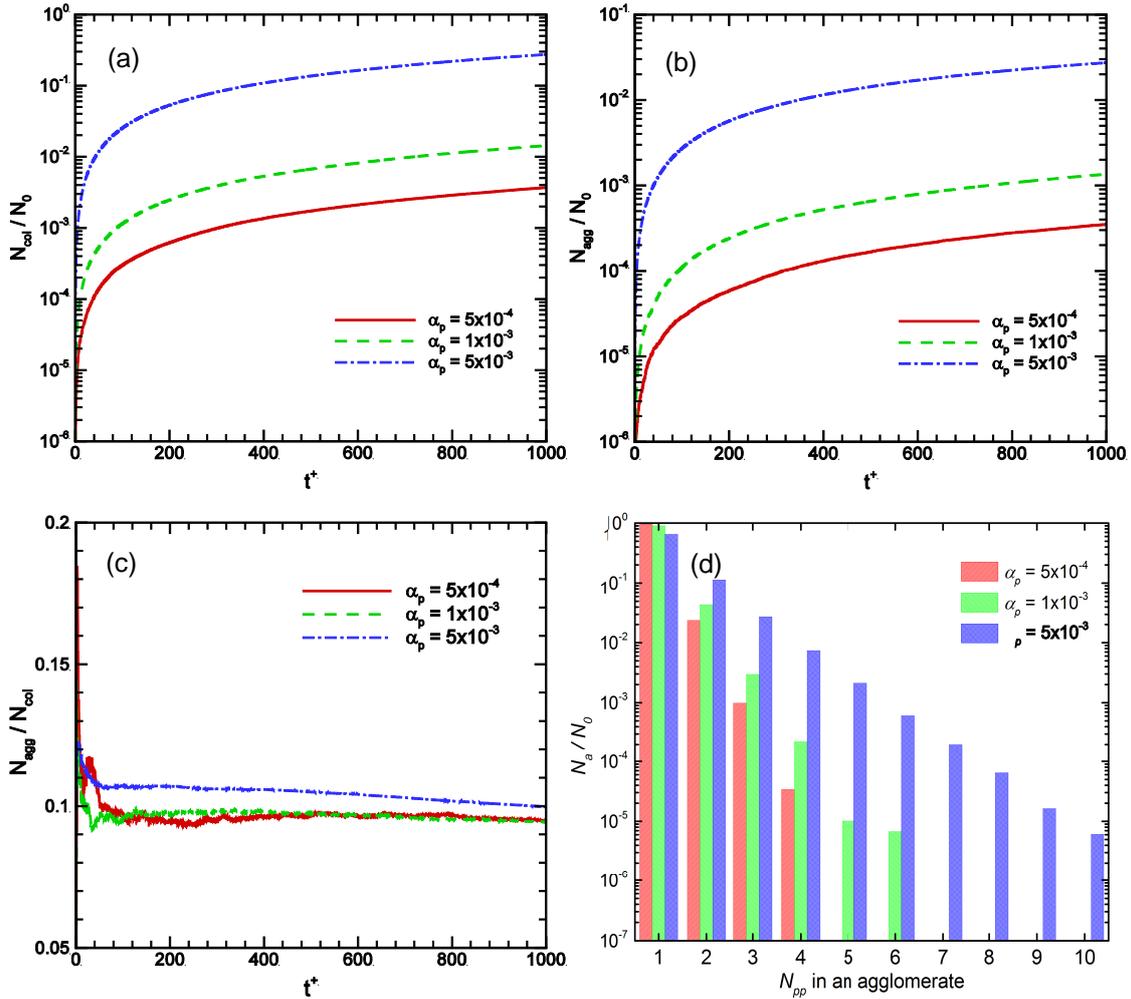

Fig. 10. Distribution of: (a) total number of particle-particle collision events, $N_{col}$, (b) total number of particle-particle collisions leading to agglomeration, $N_{agg}$, both normalised by initial total number of primary particles, $N_0$, (c) agglomeration rate, $N_{agg}/N_{col}$, all as a function of non-dimensional simulation time, $t^+$, and (d) number of particles or agglomerates, $N_a/N_0$, of the same type (single (1), double (2), triple (3), quadruple (4) etc. particles) after a simulation time, $t^+ = 1000$ ($Re_\tau = 150$, particle Stokes number $\tau_p^+ = 1$, $e_n = 0.4$, $\alpha_p = 5\times 10^{-4}$, $1\times 10^{-3}$ and $5\times 10^{-3}$).

Table 4. Values of particle-particle collisions ($N_{col}/N_0$), agglomeration events ($N_{agg}/N_0$), agglomeration rate ($N_{agg}/N_{col}$), agglomerated primary particles ($N_{app}/N_0$), agglomerate number ($N_a/N_0$) and average number of primary particles included in an agglomerate ($N_{pp}/N_a$) using different fractal dimensions after a dimensionless time $t^+ = 1000$ ($Re_\tau = 150$, $d_p = 60$ μm, $\alpha_p = 1\times 10^{-3}$).

| $\alpha_p$ | $N_{col}/N_0$ | $N_{agg}/N_0$ | $N_{agg}/N_{col}$ | $N_{pp}/N_0$ | $N_a/N_0$ | $N_{pp}/N_a$ |
|---|---|---|---|---|---|---|
| $5\times 10^{-4}$ | 0.274 | 0.026 | 0.095 | 0.050 | 0.025 | 2.042 |
| $1\times 10^{-3}$ | 0.634 | 0.184 | 0.290 | 0.307 | 0.124 | 2.488 |
| $5\times 10^{-3}$ | 0.842 | 0.154 | 0.183 | 0.265 | 0.111 | 2.382 |

## IV. CONCLUSIONS AND OUTLOOK

In this paper, an efficient and detailed CFD model has been described and used to advance our understanding of particle-particle interactions in turbulent flows, including particle agglomeration, augmenting the limited amount of work performed in this area to date. Current gaps in our understanding have arisen due to the complexity in accurately accounting for the interplay between hydrodynamics and physicochemical interactions, and the attendant computational cost involved in resolving the wide range of spatial and temporal scales involved. The developed method combines an Eulerian-Lagrangian technique, and deterministic hard sphere collision and energy balance agglomeration models, in the context of large eddy simulation. The LES, with dynamic modelling of the sub-grid scale and high nodal resolutions, provides an accurate description of a turbulent flow field that is relevant to industrial applications. The predicted accuracy of the flow field is critical in allowing proper simulation of particle transport, the interactions between particles, and between particles and solid surfaces in such flows. The predicted flow field has been found to be in very good agreement with single-phase flow results obtained from direct numerical simulations of $Re_\tau = 150$, and higher $Re_\tau$, channel flows.

The present contribution has focused on the dependency of particle-particle interactions (collision and agglomeration events) in a turbulent channel flow on the particle normal restitution coefficient, $e_n$, particle size (diameter, $d_p$, and Stokes number, $\tau_p^+$), flow shear Reynolds number, $Re_\tau$, and particle volume fraction, $\alpha_p$. In all the simulations, the particle density was kept constant at a value of 2710 kg m$^{-3}$, representative of calcite, a nuclear waste simulant. Although the results of the simulations studied in this work lack any direct quantitative comparison with physical measurements, they do lead to qualitative explanations and insights into the inter-particle collision and agglomeration processes occurring in a wall-bounded turbulent flow.

These results show the importance of the restitution coefficient to the dynamics of particle-particle collision and agglomeration in a turbulent flow. The predictions demonstrated that the normalised number of particle collisions, $N_{col}/N_0$, and the agglomeration process, $N_{agg}/N_0$, vary linearly with time, $t^+$, and are strongly dependent on the particle normal restitution coefficient, $e_n$. The larger the coefficient of restitution, the larger the number of collisions ($N_{col}/N_0$), the smaller the number of such collisions resulting in agglomeration ($N_{agg}/N_0$), and the smaller the agglomeration rate ($N_{agg}/N_{col}$), supporting previous observations [12, 24, 25]. The normal coefficient of restitution in the particle-particle interaction model controls how much of the colliding particles' kinetic energy remains for the particles to rebound from one another, compared with how much is dissipated as heat or work done in deforming the colliding pair. Hence, an increase in $e_n$ increases the amount of energy dissipated during the collision, thus reducing the probability of agglomeration conditions being satisfied.

Particle-particle interactions are also sensitive to the primary particle size (characterised by its diameter, $d_p$) or the particle inertia (characterised by its Stokes number, $\tau_p^+$), the flow inertia (characterised by the flow shear Reynolds number, $Re_\tau$) and the particle concentration (characterised by the particle volume fraction, $\alpha_p$). The relationship between particle size and particle-particle interactions shows an inverse proportionality. By analysing the particle-particle interaction events of primary particles with different sizes, it is concluded that the normalised collision frequency, $N_{col}/N_0$, collision efficiency, $N_{agg}/N_0$, and the agglomeration rate, $N_{agg}/N_{col}$, all decrease with an increase in particle size, $d_p$. Larger particles possess greater kinetic energy and thus for the same impact velocity the adhesion energy is proportionally smaller, thus facilitating the separation of the particles after a collision. On the other hand, small particles have a large surface area compared to their volume, which favours agglomeration. Furthermore, the collision of small particles was found to be the primary mechanism for particle agglomeration. Collisions among small particles lead to the formation of agglomerates much more easily than collisions involving larger, agglomerated particles. Again, this effect of particle size on particle-particle interactions is consistent with, and complements, that reported elsewhere [2, 3, 21]. Increasing the flow shear Reynolds number, $Re_\tau$, from 150 to 590 showed an increase in the normalised number of particle-particle collisions, $N_{col}/N_0$, and an increase in the normalised number of collisions leading to agglomeration, $N_{agg}/N_0$, whereas the agglomeration rate, $N_{agg}/N_{col}$, showed an inverse proportionality with, and stronger dependency on, the shear Reynolds number. An increase in the particle volume fraction, $\alpha_p$, from $\alpha_p = 5 \times 10^{-4}$ to $\alpha_p = 5 \times 10^{-3}$ decreases the space between two particles within a flow and their time between collisions, which effectively enhances particle collision and agglomeration. Overall, the sensitivity of particle-particle interaction events to the selected simulation parameters subsequently influenced the population and distribution of the primary single particles and multiple particle agglomerates formed.

The deterministic agglomeration model formulated in this work is relatively simple as it allows only binary collisions, a constant coefficient of restitution irrespective of the collision velocity, and zero friction. Other assumptions include the fact that the agglomerate structure is based on an equivalent volume sphere diameter, with the agglomerate density equal to the density of the primary particles, and with agglomeration due to van der Waals' interactions alone, neglecting any repulsive forces. However, the model adopted possesses the qualitative, and to a lesser degree, the quantitative features required to explain the experimentally observed behaviour of agglomeration processes. Some of these assumptions will be relaxed in future work to improve the capability of the model and its closeness to physical reality. The simulations have shown that there is a strong sensitivity to the normal coefficient of restitution; hence, the assumption of a constant value during agglomeration, irrespective of the collision velocity of the two particles, of whatever size, may potentially be a source of error. The assumption of an equivalent volume sphere diameter, and a constant density for all particles, are other potential areas for improvement as in reality the agglomerates will adopt fractal dimension structures and the density of the agglomerate will change depending on the packing of the primary particles. All these parameters will, in turn, change as the agglomeration process proceeds with time.

## ACKNOWLEDGMENTS

The authors would like to thank Innovate UK (formerly the Technology Strategy Board), the Engineering and Physical Sciences Research Council and the Nuclear

Decommissioning Authority for their financial support of the work described under grant 167248, "Measurement and Modelling of Sludge Transport and Separation Processes". Special thanks are also given to our industrial partners, MMI Engineering and Sellafield Ltd., for fruitful discussions during the project. Part of this work was undertaken on ARC1 and ARC2, part of the high performance computing facilities at the University of Leeds.

**REFERENCES**

1. S. Elghobashi, and G. C. Truesdell, "On the two-way interaction between homogeneous turbulence and dispersed solid particles. I: Turbulence modification," Physics of Fluids **5**, 1790 (1993).
2. C. A. Ho, and M. Sommerfeld, "Modelling of micro-particle agglomeration in turbulent flows," Chemical Engineering Science **57**, 3073 (2002).
3. B. Balakin, A. C. Hoffmann, and P. Kosinski, "The collision efficiency in a shear flow," Chemical Engineering Science **68**, 305 (2012).
4. M. U. Babler, L. Biferale, L. Brandt, U. Feudel, K. Guseva, A. S. Lanotte, C. Marchioli, F. Picano, G. Sardina, A. Soldati, and F. Toschi, "Numerical simulations of aggregate breakup in bounded and unbounded turbulent flows," Journal of Fluid Mechanics **766**, 104 (2015).
5. C. Henry, J.-P. Minier, M. Mohaupt, C. Profeta, J. Pozorski, and A. Tanière, "A stochastic approach for the simulation of collisions between colloidal particles at large time steps," International Journal of Multiphase Flow **61**, 94 (2014).
6. C. Henry, J.-P. Minier, J. Pozorski, and G. Lefèvre, "A new stochastic approach for the simulation of agglomeration between colloidal particles," Langmuir **29**, 13694 (2013).
7. C. Henry, J.-P. Minier, and G. Lefèvre, "Towards a description of particulate fouling: From single particle deposition to clogging," Advances in Colloid and Interface Science **185–186**, 34 (2012).
8. D. O. Njobuenwu, M. Fairweather, and J. Yao, "Coupled RANS–LPT modelling of dilute, particle-laden flow in a duct with a 90° bend," International Journal of Multiphase Flow **50**, 71 (2013).
9. P. A. Cundall, and O. D. L. Strack, "A discrete numerical model for granular assemblies," Géotechnique **29**, 47 (1979).
10. Y. Tsuji, T. Kawaguchi, and T. Tanaka, "Discrete particle simulation of two-dimensional fluidized bed," Powder Technology **77**, 79 (1993).
11. M. Afkhami, A. Hassanpour, M. Fairweather, and D. O. Njobuenwu, "Fully coupled LES-DEM of particle interaction and agglomeration in a turbulent channel flow," Computers & Chemical Engineering **78**, 24 (2015).
12. R. Wilson, D. Dini, and B. van Wachem, "A numerical study exploring the effect of particle properties on the fluidization of adhesive particles," AIChE Journal **62**, 1467 (2016).
13. C. T. Crowe, J. D. Schwarzkopf, M. Sommerfeld, and Y. Tsuji, *Multiphase flows with droplets and particles* (CRC press, 2011).
14. B. P. B. Hoomans, J. A. M. Kuipers, W. J. Briels, and W. P. M. van Swaaij, "Discrete particle simulation of bubble and slug formation in a two-dimensional gas-fluidised bed: A hard-sphere approach," Chemical Engineering Science **51**, 99 (1996).
15. M. Chen, K. Kontomaris, and J. B. McLaughlin, "Direct numerical simulation of droplet collisions in a turbulent channel flow. Part I: collision algorithm," International Journal of Multiphase Flow **24**, 1079 (1999).

16. S. Sundaram, and L. R. Collins, "Numerical considerations in simulating a turbulent suspension of finite-volume particles," Journal of Computational Physics **124**, 337 (1996).
17. P. Kosinski, and A. C. Hoffmann, "Extended hard-sphere model and collisions of cohesive particles," Physical Review E **84**, 031303 (2011).
18. Y. Yamamoto, M. Potthoff, T. Tanaka, T. Kajishima, and Y. Tsuji, "Large-eddy simulation of turbulent gas–particle flow in a vertical channel: effect of considering inter-particle collisions," Journal of Fluid Mechanics **442**, 303 (2001).
19. M. Sommerfeld, "Validation of a stochastic Lagrangian modelling approach for inter-particle collisions in homogeneous isotropic turbulence," International Journal of Multiphase Flow **27**, 1829 (2001).
20. N. G. Deen, M. Van Sint Annaland, M. A. Van der Hoef, and J. A. M. Kuipers, "Review of discrete particle modeling of fluidized beds," Chemical Engineering Science **62**, 28 (2007).
21. J. Wang, Q. Shi, Z. Huang, Y. Gu, L. Musango, and Y. Yang, "Experimental investigation of particle size effect on agglomeration behaviors in gas–solid fluidized beds," Industrial & Engineering Chemistry Research **54**, 12177 (2015).
22. D. Jürgens, "Modellierung und Simulation der Partikelagglomeration in turbulenten,dispersen Mehrphasenströmungen," Master's Thesis Helmut-Schmidt-Universität, 2012.
23. M. Alletto, "Numerical investigation of the influence of particle–particle and particle–wall collisions in turbulent wall-bounded flows at high mass loadings," Ph.D. Thesis Helmut-Schmidt University, 2014.
24. M. Breuer, and N. Almohammed, "Modeling and simulation of particle agglomeration in turbulent flows using a hard-sphere model with deterministic collision detection and enhanced structure models," International Journal of Multiphase Flow **73**, 171 (2015).
25. N. Almohammed, and M. Breuer, "Modeling and simulation of agglomeration in turbulent particle-laden flows: A comparison between energy-based and momentum-based agglomeration models," Powder Technology **294**, 373 (2016).
26. P. Kosinski, and A. C. Hoffmann, "An extension of the hard-sphere particle–particle collision model to study agglomeration," Chemical Engineering Science **65**, 3231 (2010).
27. W. M. Weber, K. D. Hoffman, and M. C. Hrenya, "Discrete-particle simulations of cohesive granular flow using a square-well potential," Granular Matter **6**, 239 (2004).
28. B. Derjaguin, and L. Landau, "Theory of the stability of strongly charged lyophobic sols and of the adhesion of strongly charged particles in solutions of electrolytes," Acta Physico Chemica URSS **14**, 633 (1941).
29. E. J. W. Verwey, and J. T. G. Overbeek, *Theory of the stability of lyophobic colloids* (Elsevier, 1948).
30. F. Thielmann, M. Naderi, M. A. Ansari, and F. Stepanek, "The effect of primary particle surface energy on agglomeration rate in fluidised bed wet granulation," Powder Technology **181**, 160 (2008).
31. M. Alletto, and M. Breuer, "One-way, two-way and four-way coupled LES predictions of a particle-laden turbulent flow at high mass loading downstream of a confined bluff body," International Journal of Multiphase Flow **45**, 70 (2012).
32. M. Breuer, and M. Alletto, "Efficient simulation of particle-laden turbulent flows with high mass loadings using LES," International Journal of Heat and Fluid Flow **35**, 2 (2012).
33. D. O. Njobuenwu, and M. Fairweather, "Deterministic modelling of particle agglomeration in turbulent flow,"Proceedings of the Eighth International Symposium on Turbulence, Heat and Mass Transfer, K. Hanjalic, T. Miyauchi, D. Borello, M. Hadziabdic and P. Venturini (Eds.), Begell House Inc., New York, 587, 2015.

34. S. Matsumoto, and S. Saito, "Monte Carlo simulation of horizontal pneumatic conveying based on the rough wall model," Journal of Chemical Engineering of Japan **3**, 223 (1970).
35. M. Bini, and W. P. Jones, "Large-eddy simulation of particle-laden turbulent flows," Journal of Fluid Mechanics **614**, 207 (2008).
36. D. O. Njobuenwu, and M. Fairweather, "Simulation of inertial fibre orientation in turbulent flow," Physics of Fluids **28**, 063307 (2016).
37. U. Piomelli, and J. Liu, "Large-eddy simulation of rotating channel flows using a localized dynamic model," Physics of Fluids **7**, 839 (1995).
38. R. Mei, "An approximate expression for the shear lift force on a spherical particle at finite reynolds number," International Journal of Multiphase Flow **18**, 145 (1992).
39. J. M. Tingey, B. C. Bunker, G. L. Graff, K. D. Keeper, A. S. Lea, and D. R. Rector, "Colloidal agglomerates in tank sludge and their impact on waste processing," MRS Online Proceedings Library Archive **556**, 1315 (1999).
40. W. J. Stronge, *Impact mechanics* (Cambridge university press, 2004).
41. G. K. El Khoury, P. Schlatter, A. Noorani, P. F. Fischer, G. Brethouwer, and A. V. Johansson, "Direct numerical simulation of turbulent pipe flow at moderately high Reynolds numbers," Flow, Turbulence and Combustion **91**, 475 (2013).
42. C. Marchioli, A. Soldati, J. G. M. Kuerten, B. Arcen, A. Tanière, G. Goldensoph, K. D. Squires, M. F. Cargnelutti, and L. M. Portela, "Statistics of particle dispersion in direct numerical simulations of wall-bounded turbulence: Results of an international collaborative benchmark test," International Journal of Multiphase Flow **34**, 879 (2008).
43. L.-P. Wang, A. S. Wexler, and Y. Zhou, "Statistical mechanical description and modelling of turbulent collision of inertial particles," Journal of Fluid Mechanics **415**, 117 (2000).
44. C.-L. Lin, and M.-Y. Wey, "The effect of mineral compositions of waste and operating conditions on particle agglomeration/defluidization during incineration," Fuel **83**, 2335 (2004).
45. G. V. Reddy, and S. K. Mahapatra, "Effect of coal particle size distribution on agglomerate formation in a fluidized bed combustor (FBC)," Energy Conversion and Management **40**, 447 (1999).
46. M. Dosta, S. Antonyuk, and S. Heinrich, "Multiscale simulation of agglomerate breakage in fluidized beds," Industrial & Engineering Chemistry Research **52**, 11275 (2013).
47. M. Ernst, and M. Sommerfeld, "On the volume fraction fffects of inertial colliding particles in homogeneous isotropic turbulence," Journal of Fluids Engineering **134**, 031302 (2012).